 \documentclass{elsart}

\usepackage{icarus}
\usepackage{lineno}

\usepackage{natbib}

\bibpunct{(}{)}{;}{a}{,}{,}

\usepackage{graphicx}

\newcommand{\BibTeX}{ \textrm{B\kern-.05em\textsc{i\kern-.025em b}\kern-.08em
    T\kern-.1667em\lower.7ex\hbox{E}\kern-.125emX} }

\begin{document}

\begin{frontmatter}



\title{1D photochemical model of the ionosphere and the stratosphere of Neptune}


\author[LAB]{M. Dobrijevic} 
\author[ISM]{J.C. Loison}
\author[SWRI]{V. Hue}
\author[LAB,LESIA]{T. Cavali\'e} 
\author[ISM]{K.M. Hickson}

\address[LAB]{Laboratoire d'Astrophysique de Bordeaux, Univ. Bordeaux, CNRS, B18N, all\'ee Geoffroy Saint-Hilaire, 33615 Pessac, France}
\address[ISM]{Institut des Sciences Mol\'eculaires (ISM), CNRS, Univ. Bordeaux, 351 cours de la Lib\'eration, 33400 Talence}
\address[SWRI]{Southwest Research Institute, San Antonio, TX 78228, United States.}
\address[LESIA]{LESIA, Observatoire de Paris, Universit\'e PSL, CNRS, Sorbonne Universit\'e, Univ. Paris Diderot, Sorbonne Paris Cit\'e, 5 place Jules Janssen, 92195 Meudon, France}



%
%
%
%
%


\end{frontmatter}



\begin{flushleft}
\vspace{1cm}
Number of pages: \pageref{lastpage} \\
Number of tables: \ref{lasttable}\\
Number of figures: \ref{lastfig}\\
\end{flushleft}


\begin{pagetwo}{Photochemistry of Neptune's atmosphere}

Michel Dobrijevic \\
Laboratoire d'Astrophysique de Bordeaux \\
2 rue de l'observatoire, Floirac, F-33271, France. \\
\\
Email: Michel.Dobrijevic$@$obs.u-bordeaux1.fr\\
Phone: +33-5-5777-6124 \\

\end{pagetwo}

\begin{abstract}
Neptune remains a mysterious world that deserves further exploration and is a high-priority objective for a future planetary mission in order to better understand the formation and evolution of ice giant planets.
We have developed a coupled ion-neutral 1D photochemical model of Neptune's atmosphere to study the origin and evolution of the hydrocarbons and the oxygen species. The up-to-date chemical scheme is derived from one used for Titan's atmosphere, which led to good agreements with the Cassini-CIRS observations for oxygen species and the main hydrocarbons.
The main results we obtain are the following: The ion-neutral chemistry coupling produces aromatics (and benzene in particular) in the atmosphere of Neptune with relatively high abundances. Our model results are in good agreement with observations (taking model  uncertainties into account). Two ionospheric peaks are present in the atmosphere located above the pressure level of 10$^{-5}$ mbar and around 10$^{-3}$ mbar. The influx of oxygen species in the upper atmosphere of Neptune has an effect on the concentration of many ions. 
We show that in situ exploration of Neptune's atmosphere would provide very interesting constraints for photochemical models concerning in particular the origin of oxygen species and the contribution of ion chemistry. A precise description of upper atmospheric chemistry is crucial for a better understanding of the internal composition and the formation processes of this planet.
\end{abstract}

\begin{keyword}
Neptune \sep Photochemistry \sep Atmospheres \sep Ionospheres \sep aromatics
\end{keyword}


\section{Introduction}
\label{section:introduction}

Neptune, the outermost planet of the Solar System, is an ice giant that has only been visited by the Voyager 2 spacecraft 30 years ago. Its atmosphere is composed mainly of molecular hydrogen (H$_2$) and helium (He), with a small fraction of methane CH$_4$. Solar photoionization of these species generates an ionosphere whose composition is unknown. The photodissociation of CH$_4$ produces radicals which react to produce hydrocarbons. Only a few of them have been detected so far (CH$_3$ \citep{Bezard1999}, C$_2$H$_2$, C$_2$H$_4$, C$_2$H$_6$ \citep{Fletcher2010,Greathouse2011}, CH$_3$C$_2$H and C$_4$H$_2$ (\citep{Meadows2008}). Other species (H$_2$O, CO, CO$_2$ \citep{Feuchtgruber1997,Meadows2008,Lellouch2005,Luszcz-Cook2013}, HCN \citep{Rezac2014}, and CS \citep{Moreno2017}) have also been detected and their origin is still debated. The ionosphere and stratosphere of this distant planet are not well characterized because remote observations from Earth-based facilities are challenging and also because of the lack of exploration missions.

Benzene has been detected in Jupiter and Saturn \citep{Kim1985, Bezard2001, Guerlet2015} but not in Uranus and Neptune (see for instance \citet{Bezard2001} for a discussion on that point). This fact might indicate that the photochemistry of hydrocarbons in Neptune is less efficient than in Jupiter and Saturn. Indeed, 1D neutral photochemical models of hydrocarbons do not predict C$_6$H$_6$ to be abundant in Neptune's atmosphere \citep{Moses2005}. However, photochemical models of Titan \citep{Vuitton2018, Loison2018} show that the ion chemistry favors the production of C$_6$H$_6$, suggesting that C$_6$H$_6$ might also be efficiently produced in the atmosphere of Neptune. This motivates the need to develop a coupled ion-neutral photochemical model for this planet.

Three oxygen-bearing species, carbon monoxide (CO), water (H$_2$O) and carbon dioxide (CO$_2$), have been detected in the stratosphere of Neptune and are supplied by one or several external sources. While H$_2$O and CO$_2$ seem to be delivered by interplanetary dust particles \citep{Moses2017}, CO (along with HCN and CS) was probably delivered by a comet a few centuries ago \citep{Lellouch2005, Lellouch2010, Moreno2017}. We note that CO also has an internal source \citep{Lellouch2005}, whose magnitude is still debated \citep{Luszcz-Cook2013, Teanby2019}. A striking particularity is that the relative abundance of Neptune's stratospheric CO is about two orders of magnitude greater than in any other giant planet of the Solar System. According to \citet{Moses2017}, it is expected that the influx of oxygen species affects the chemistry of both the ionosphere and the stratosphere of giant planets. In particular, they have pointed out that the large abundance of CO has an effect on the production of hydrocarbons, leading in particular to a decrease of C$_2$H$_2$ and higher-order hydrocarbons. However, they noted that this result could be an artifact due to the use of the low-resolution absorption cross section they use for CO. This point highlights the need to develop an ion-neutral chemical model with high-resolution cross sections that also accounts for the influx of oxygen species. 

The physical and chemical properties of Neptune's ionosphere are poorly known, due to the sparse number of observations available. A narrow ionization layer has been detected by Voyager 2 around 1400 km of altitude \citep{Tyler1989}, where the zero of altitude corresponds to the 1-bar pressure level. From an analysis of radio-occultation data acquired with Voyager 2, \citet{Lindal1992} obtained a layer between 500 km ($6\times10^{-4}$ mbar) and 1000 km ($1.3\times10^{-6}$ mbar) with an electron concentration reaching about $2\times10^4$ cm$^{-3}$ around 800 km ($8\times10^{-3}$ mbar). From previous ionospheric models and comparison with other giant planets, it is expected that H$^+$ and H$_3^+$ are the dominant ions in the ionosphere of Neptune \citep{Atreya1984, Lyons1995}. 

Recently, \citet{Arridge2014} and \citet{Turrini2014} discussed the importance of dedicated missions to explore the Ice Giants, with \citet{Mousis2018} highlighting the benefits of in situ exploration, in order to determine their physical condition, atmospheric and internal compositions. An up-to-date photochemical model accounting for coupled ion-neutral chemistry is required to improve the predictions of the atmospheric species in the context of such potential missions with in situ probes. 

In section 2, we present the photochemical model. The results regarding the hydrocarbons, the oxygen species and the main ions are presented in section 3. Some results are discussed in section 4 and we conclude in section 5.

\section{Model}
\label{section:model}

The 1D photochemical model used in the present study is similar to the recent model developed for Titan, which couples the ions and the neutral species of hydrocarbons, oxygen and nitrogen species \citep{Dobrijevic2016}. In the following section, only the major modifications or additions, as well as Neptune's specific features are outlined.

\subsection{Atmospheric model}

\begin{figure}[htp]
\begin{center}
\includegraphics[width=1.0\columnwidth]{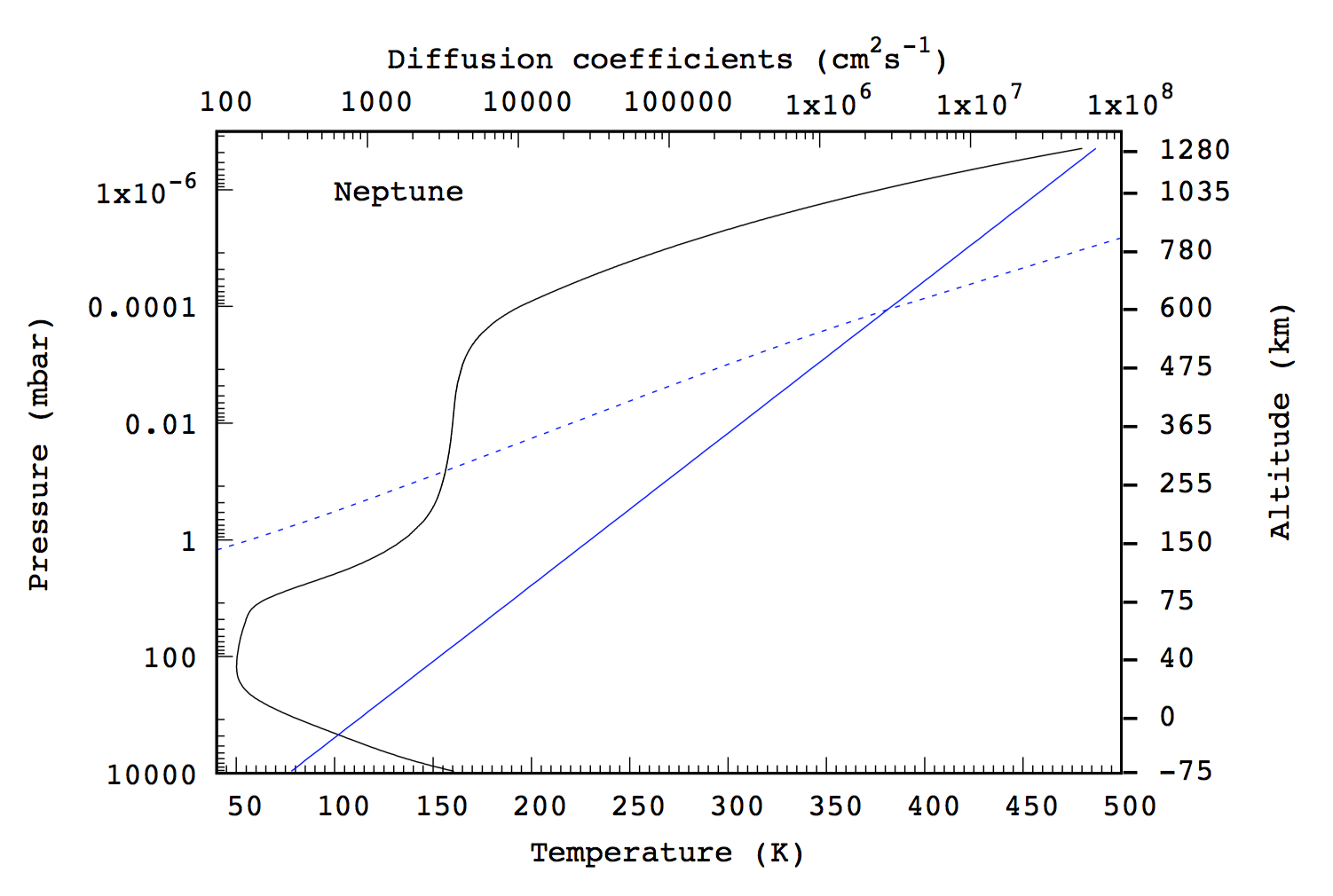} 
\caption{Black curve and lower scale: Temperature profile as a function of pressure (adapted from \citet{Feuchtgruber2013}). Blue curves and upper scale: Profiles of the eddy diffusion coefficient (solid line) and CH$_4$ molecular diffusion coefficient (dashed-line) adopted in the present model.}
\label{fig:eddy}
\end{center}
\end{figure}

The pressure-temperature profile used in the present study, shown in Figure \ref{fig:eddy}, is adapted from the temperature profile derived by \citet{Feuchtgruber2013} from Herschel data. This profile is slightly different from the one derived earlier by \citet{Fletcher2010} from Akari data. Different profiles have been inferred from various observations, resulting in an uncertainty of about 5 K around the tropopause and 10 K in the stratosphere. Such differences would not have a significant effect on the model results for Neptune (compared to uncertainties due to the chemical rate constants) as shown by \citet{Dobrijevic2016b} for Titan. In order to compute the altitude grid, we use the physical parameters of Neptune at the equator.

\subsection{Boundary conditions}

The presence of CO in the stratosphere of Neptune is likely due to a cometary impact (see \citet{Moses2017} for a recent discussion on that subject). Since the main focus of the present study is not to study the origin and evolution of oxygen species, we follow the methodology used by \citet{Moses2018}, which provides a reasonable approximation. We have assumed, in the first place, that H$_2$O, CO and CO$_2$ are supplied by interplanetary dust particles (IDP) with influx rates $\Phi_i$ at the top boundary similar to \citet{Moses2018} ($\Phi_{H_2O} = 2\times10^5$ cm$^{-2}$s$^{-1}$, $\Phi_{CO} = 10^3\times \Phi_{H_2O}$, $\Phi_{CO_2} = 0.15\times\Phi_{H_2O}$). We have also tested a scenario based on a cometary impact to explain the origin of CO and study a putative effect of ionic chemistry. Indeed, \citet{Lellouch2005} proposed a dual origin for CO in the atmosphere of Neptune with stratospheric CO resulting from a large cometary impact that occurred $\approx$ 200 years ago. Following \citet{Moses2017}, we have also tested this hypothesis in our ion-neutral model and we have considered a sudden cometary supply of CO with an initial mole fraction of 10$^{-4}$ above 0.1 mbar. The background influxes of H$_2$O and CO$_2$ were similar to the ones used in the IDP model and the constant influx of CO at the top of the atmosphere was set to zero. 

Unlike previous photochemical models \citep{Moses2005, Dobrijevic2010, Moses2018}, we do not include a downward flux of atomic hydrogen at the upper boundary to account for additional photochemical production in the high atmosphere. In the present study, we assume that photoionization and subsequent ionic chemistry are responsible for this source previously added to the model. All other species are assumed to have zero-flux boundary conditions at the top of the model atmosphere ($2.0\:10^{-7}$ mbar). 

At the lower boundary (1 bar), we set the mole fraction of He, CH$_4$ and H$_2$ respectively to $y_{He} = 0.149$ \citep{Burgdorf2003}, $y_{CH_4} = 9.3\times 10^{-4}$ \citep{Lellouch2015}, $y_{CO} = 8\times 10^{-8}$ \citep{Moses2017} and $y_{H_2} = 1.0 - y_{He} - y_{CH_4} - y_{CO}$. As currently done in photochemical modeling, the mole fraction of CH$_4$ is set to the stratospheric value, which is supersaturated (greater than the value it should be considering the cold trap at the tropopause). Consequently, our photochemical model is not adapted to the troposphere of Neptune. All other compounds have a downward flux given by the maximum diffusion velocity $v = K_{zz}/H$ where $K_{zz}$ is the eddy diffusion coefficient and $H$ the atmospheric scale height at the lower boundary.

To summarize, we consider the following different models: in model A, CO is supplied by a steady external source (e.g. IDP) at the top of the atmosphere and a fixed mole fraction at the lower boundary. The integration time is stopped when the steady state is reached. Model A is separated into two submodels: model A1 concerns only neutrals, while model A2 couples ions and neutrals in order to highlight the differences brought about by ionic chemistry. In model B1, we compute the steady state with no external supply of CO (the CO mole fraction is fixed at the lower boundary) with the coupled ion-neutrals scheme, which serves as an initial atmospheric composition for model B2. In this latter submodel, CO is also supplied by a cometary impact and the evolution of species abundances is studied as a function of time in the coupled ions-neutrals model. 

Descriptions of the different models are summarized in  Table \ref{tab:models}. In the four models, the external fluxes of H$_2$O and CO$_2$ have the fixed values given above.

\begin{table}[htp]
\caption{Summary of the different models considered in the present study.}
\begin{center}
\begin{tabular}{|c|p{2.5cm}|p{2.5cm}|p{2.5cm}|p{2.5cm}|}
\hline
Type	& Internal and steady external supply of CO & Internal and steady external supply of CO & Internal supply of CO only & Comet-impact hypothesis \\
\hline
Chemistry	& neutral & ion-neutral & ion-neutral & ion-neutral \\
Steady-state & yes & yes & yes & no \\
Name of the model	& A1 & A2 & B1 & B2 \\
\hline
\end{tabular}
\end{center}
\label{tab:models}
\label{lasttable}
\end{table}

\subsection{Vertical transport}

In a 1D photochemical model, vertical transport is dominated by molecular diffusion at altitudes above the homopause and eddy mixing below this limit. The homopause for a given species is the altitude level for which the eddy diffusion coefficient $K_{zz}$ is equal to the molecular diffusion coefficient for this species. The $K_{zz}$ profile is a free parameter of 1D photochemical models. The best way to constrain this parameter is to compare the model results with observational data of some particular species. In the case of Titan, an inert species like argon (Ar) is very useful for this purpose. For the giant planets, the situation is more difficult. The homopause can be constrained using CH$_4$ observations in the upper atmosphere since its profile is driven by molecular diffusion. Below the homopause, the $K_{zz}$ profile is usually constrained by a comparison between observations and model results for the main hydrocarbons. Unfortunately, this is a very imprecise methodology since model results have strong uncertainties (see for instance \citet{Dobrijevic2010} for Neptune and \citet{Dobrijevic2011} for Saturn), which can be much larger than uncertainties on observational data and variations expected from seasonal effects (see \citet{Moses2018}). 
In the present study we use CH$_4$ observations in the upper atmosphere to constrain the value of $K_{zz}$ at the homopause. We use a simple expression for the eddy profile given by: 
\begin{equation}
K(z) = K_0*(P_0/P(z))^{0.5}
\end{equation}
where $K_0 = 3\times 10^2$ cm$^2$s$^{-1}$ and $P_0 = 10^4$ mbar have been adjusted to obtain a CH$_4$ mole fraction profile in agreement with observations. This eddy profile shown in Figure \ref{fig:eddy} also gives quite good agreement with other hydrocarbons, as shall be shown in Section \ref{section:neutral}. 

\subsection{Chemical scheme}

Only hydrocarbons and oxygen species are considered in the present study. The chemical scheme is similar to the one used for Titan in \citet{Loison2018}, in which the photochemistry of aromatics was updated, with a few additional modifications as described below. We do not consider nitrogen compounds in the present model. We introduce He$^+$ reactions, as well as H$_2$O, CO and CO$_2$ photoionizations and reactions involving H$_2$O$^+$, CO$^+$, CO$_2^+$. For the interaction of excited benzene with H$_2$, we consider that this process results only in relaxation in a similar manner to excited benzene collisions with N$_2$. It should be noted that the introduction of photoionizations for H$_2$O, CO and CO$_2$ have little influence for Neptune but their importance should be verified for the other gas giants.
The list of reactions and their column production rates (integrated over altitude) are given in Appendix \ref{appendix:rates}.

\subsection{Electron production}

We consider that the ionosphere of Neptune is only generated by solar EUV radiation, which produces photoelectrons. Details on the cross sections and branching ratios for the photoionization processes are presented in \citet{Dobrijevic2016} and \citet{Loison2018}. Other sources of electrons such as magnetospheric electrons (ME) and Galactic Cosmic Rays (GCR) are not taken into account in the present model. \citet{Bishop1995} argued that the magnetospheric power input is relatively small in the auroral region of Neptune (10$^4$ lower than for Jupiter) and large latitudinal variation in the behaviour of Neptune's ionosphere due to magnetospheric electrons might be low, in contrast to solar radiation, as shown by \citet{Galand2009} for Saturn for instance. However, these two additional sources of electrons might have an important impact on the ionic chemistry in the magnetic polar regions and in the lower atmosphere and would merit further studies. Also, electron transport and ionization by secondary electrons are not considered in the present model. \citet{Galand2009} studied secondary ionization at Saturn and found that the secondary production rate affects both the density of the ionospheric peak and the peak altitude. Hence, the inclusion of these effects would improve current models of Neptune's atmosphere.

\citet{Bishop1995} noted that, within the range of uncertainties, the ion and neutral temperatures seem similar in the ionosphere (above 2000 km of altitude). In order to compute the dissociative recombination rates of positive ions, we consider that the electron temperature is equal to the neutral temperature as a function of altitude.

\section{Results}
\label{section:results}

\subsection{Neutral hydrocarbons abundances}
\label{section:neutral}

\begin{figure}[htp]
\begin{center}
\includegraphics[width=1.0\columnwidth]{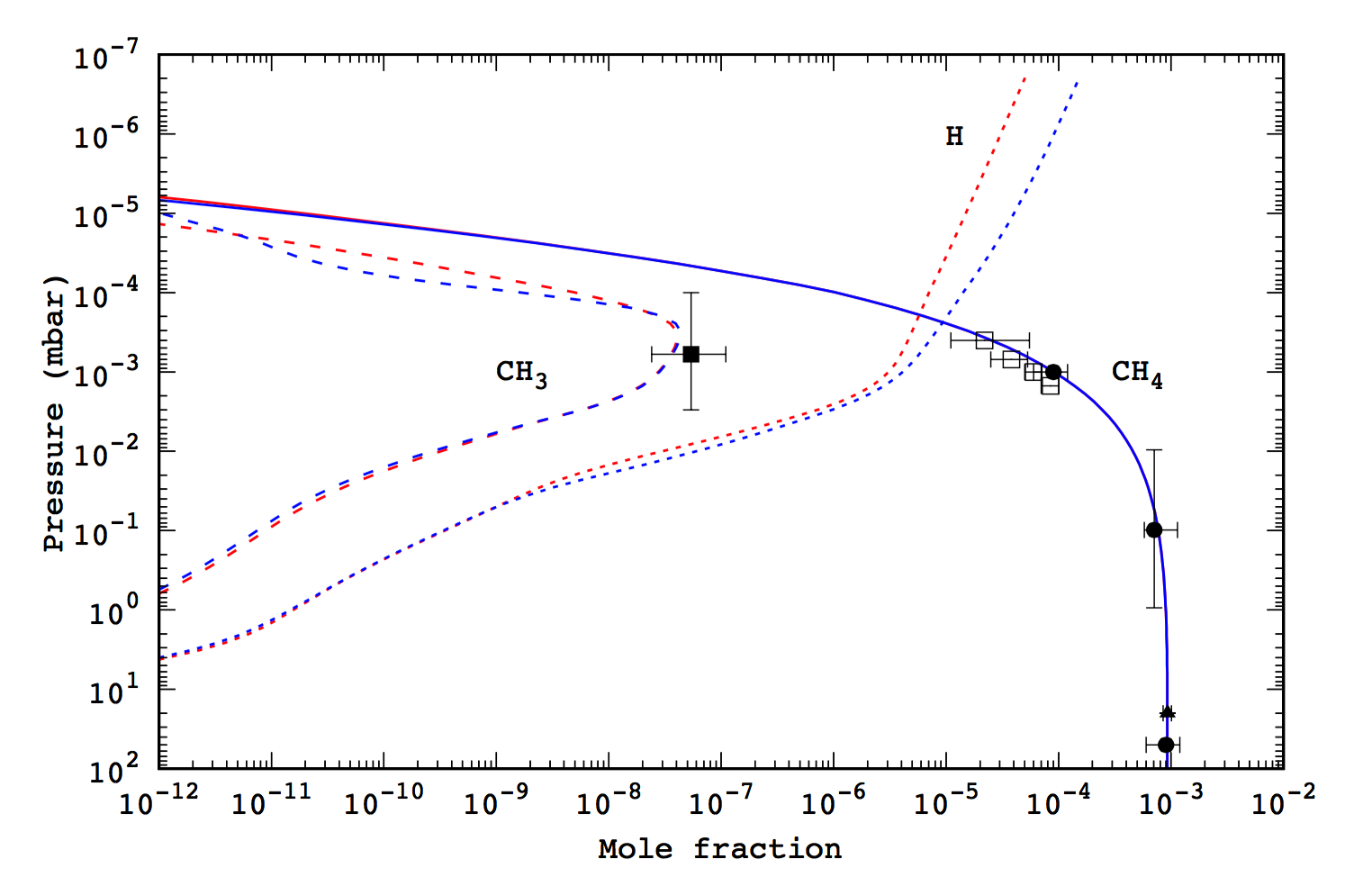} 
\caption{Mole fraction profiles of methane (CH$_4$), methyl radicals (CH$_3$) and atomic hydrogen (H) as a function of pressure derived from the neutral model (model A1, in red) and the coupled ion-neutral model (model A2, in blue). Observations are given in black: full square for CH$_3$ \citep{Bezard1999}, empty squares \citep{Yelle1993}, triangle \citep{Lellouch2015}, full circles \citep{Fletcher2010} for CH$_4$.}
\label{fig:CH4}
\end{center}
\end{figure}

\begin{figure}[htp]
\begin{center}
\includegraphics[width=1.0\columnwidth]{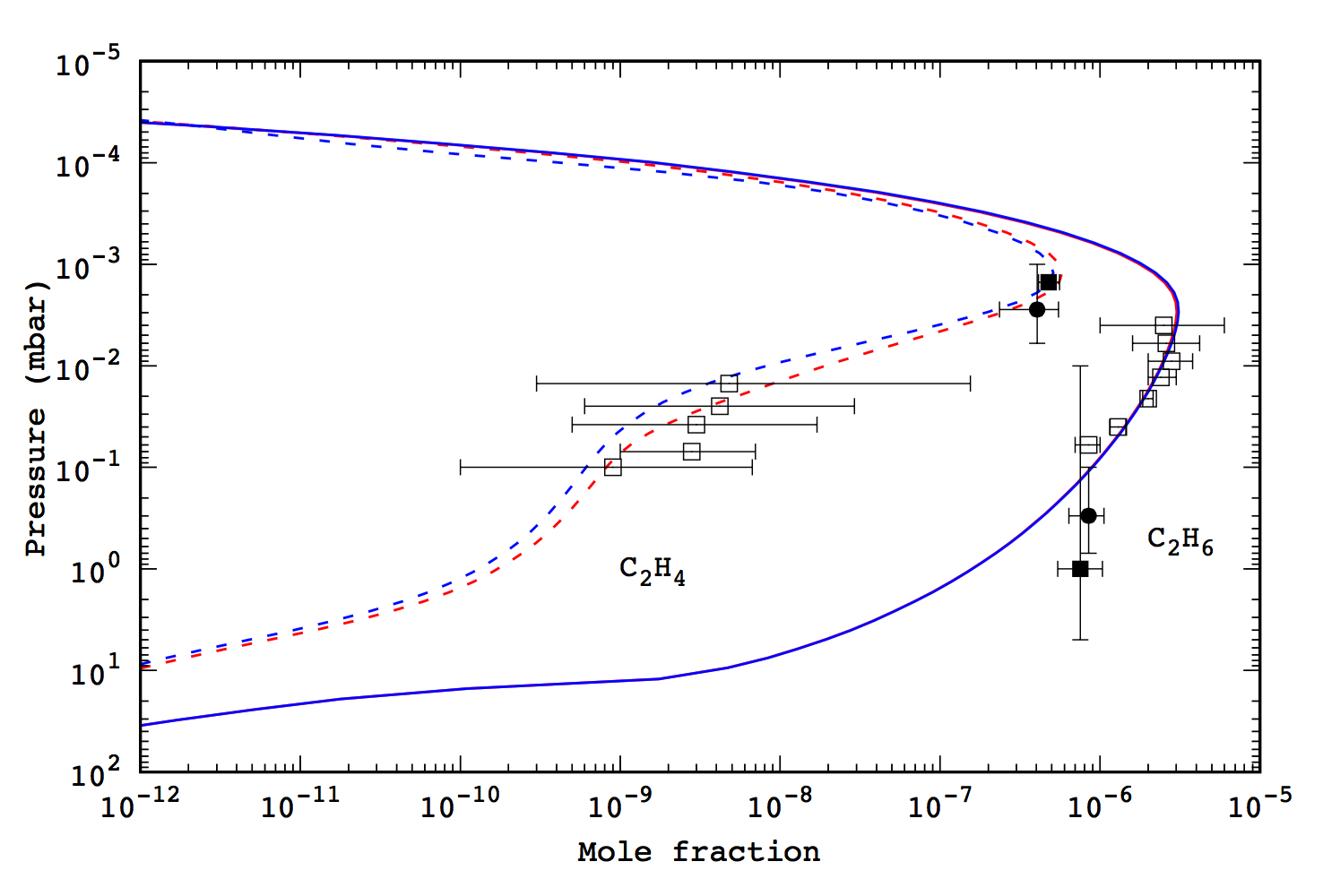} 
\includegraphics[width=1.0\columnwidth]{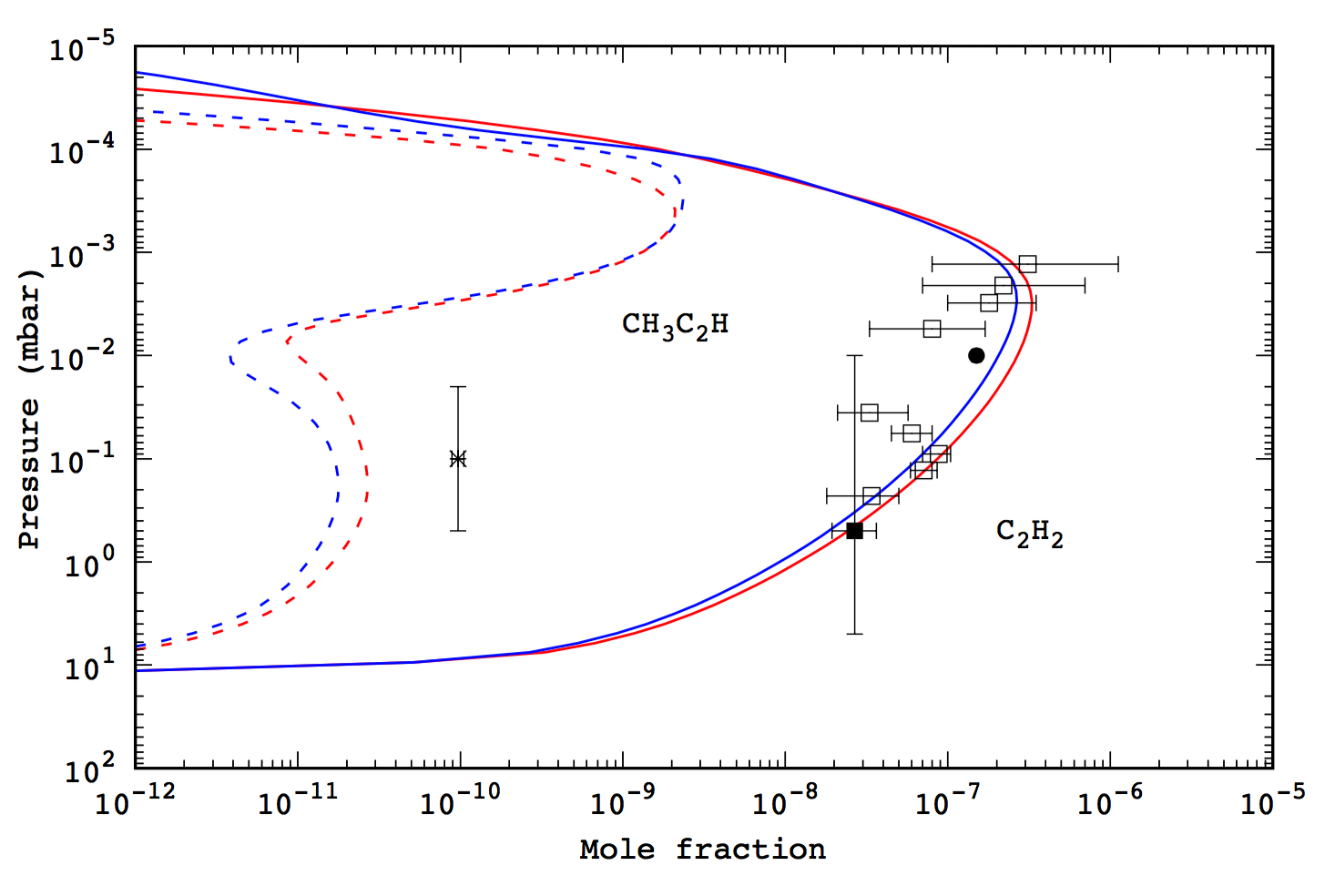} 
\caption{Top: Mole fraction profiles of ethane (C$_2$H$_6$) and ethylene (C$_2$H$_4$) as a function of pressure derived from the neutral model (model A1, in red) and the coupled ion-neutral model (model A2, in blue). Bottom:  Mole fraction profiles of acetylene (C$_2$H$_2$) and methylacetylene (CH$_3$C$_2$H). Observations are shown in black: star for CH$_3$C$_2$H \citep{Meadows2008} and full squares \citep{Greathouse2011}, empty squares \citep{Yelle1993}, full circles \citep{Fletcher2010} for C$_2$H$_2$, C$_2$H$_4$ and C$_2$H$_6$.}
\label{fig:C2}
\end{center}
\end{figure}

\begin{figure}[htp]
\begin{center}
\includegraphics[width=1.0\columnwidth]{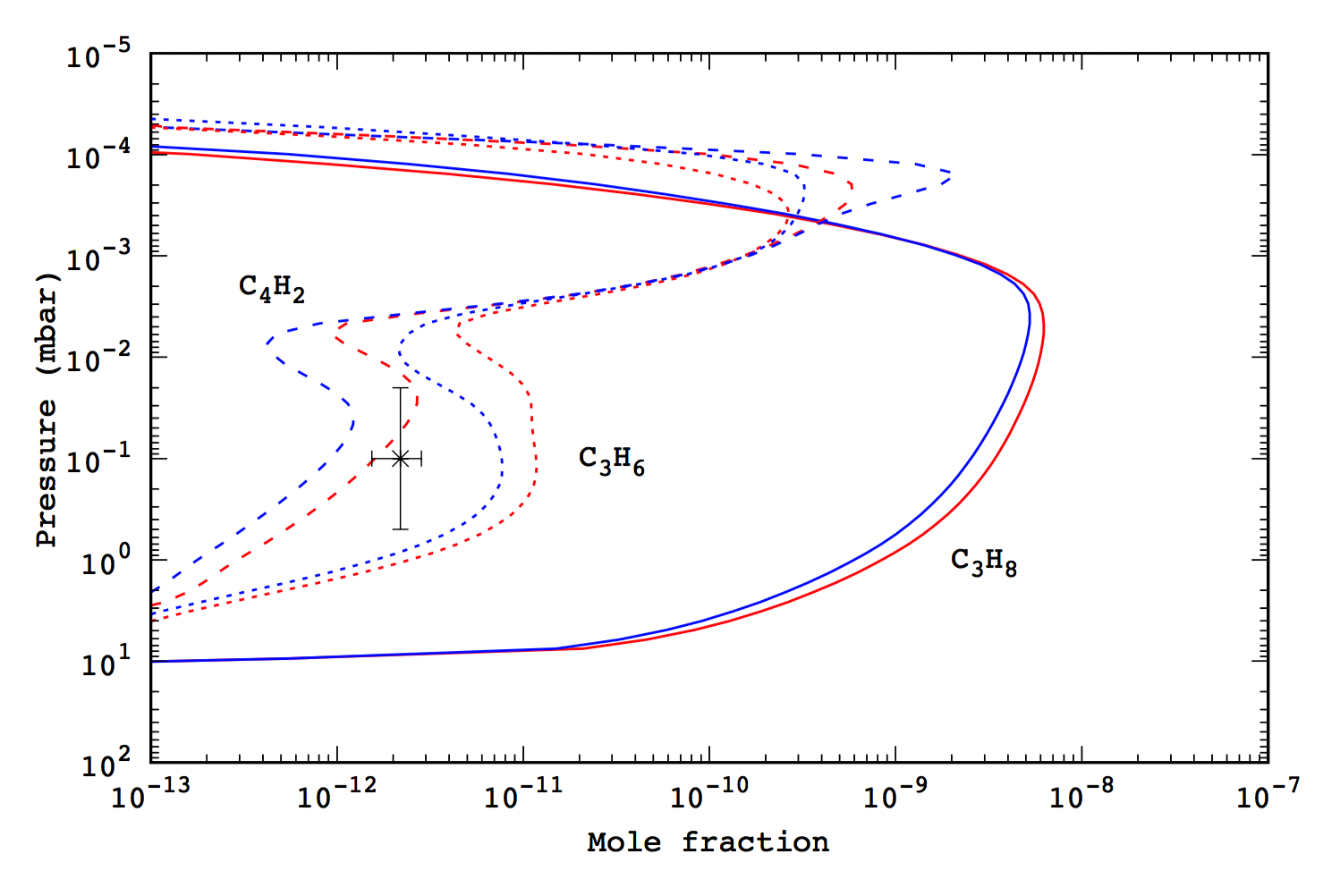} 
\caption{Mole fraction profiles of propane (C$_3$H$_8$), diacetylene (C$_4$H$_2$) and propene (C$_3$H$_6$). See previous caption. The observation of C$_4$H$_2$ is shown by a black star \citep{Meadows2008}.}
\label{fig:C3}
\end{center}
\end{figure}

Figure \ref{fig:CH4} shows the mole fraction profiles of methane (CH$_4$) and methyl radicals (CH$_3$) obtained using the eddy diffusion coefficient presented in Figure \ref{fig:eddy}. Both species are in good agreement with the handful of observations, indicating that the location of the homopause should be quite well constrained. We find that the homopause is located around $1.2\times 10^{-4}$ mbar with a $K_{zz}$ equal to $2.7\times 10^6$ cm$^2$s$^{-1}$ at this level. This eddy profile gives also a quite good agreement between model results and observations for most of the other species (considering uncertainties on observations and model results). 
It is noteworthy to recall that the comparison of model results with data obtained from various observations recorded at different times and using different temperature profiles is just a first-order approximation. To be more conclusive, it would be better to simulate all the past observations with the temperature profile and abundance profiles used in the present study, which would require a huge amount of work, and is beyond the scope of this paper.
Given this, we see in Figures \ref{fig:C2} and \ref{fig:C3} that model results are in quite good agreement with all observations (taking into account uncertainties on model results and observations), except for CH$_3$C$_2$H and C$_4$H$_2$ at first glance. However, the uncertainties on model results are quite large on these two species (see \citet{Dobrijevic2010} for Neptune and \citet{Dobrijevic2011} for Saturn). This point is discussed in section \ref{section:UP_hydroc}. The key reactions associated to these two species should be studied in priority to lower the uncertainties on model results (see \citet{Hebrard2013} for the methodology).
We also note that the column density of C$_2$H$_4$ inferred from our model A2 is equal to $3.2\times 10^{14}$ molecules cm$^{-2}$ above the 0.2-mbar level, in quite good agreement with the data obtained by \citet{Schulz1999} from ISO/PHT-S observations (between $1.1\times 10^{14}$ and $3.0\times 10^{14}$ molecules cm$^{-2}$).

Propene (C$_3$H$_6$) has not been detected yet in Neptune's atmosphere. This hydrocarbon has been detected in Titan's atmosphere \citep{Nixon2013} and the abundance predicted from the photochemical model of \citet{Loison2015} is in good agreement with this observation. The chemical scheme we use in the present study originates from this photochemical model for most of the C$_2$-C$_3$H$_x$ scheme, which makes us quite confident about the C$_3$H$_6$ profile obtained for Neptune (see Figure \ref{fig:C3}). Propene appears to be more abundant than diacetylene in our model and could be detectable.

\subsection{Aromatic abundances}
\label{section:arom}

\begin{figure}[htp]
\begin{center}
\includegraphics[width=1.0\columnwidth]{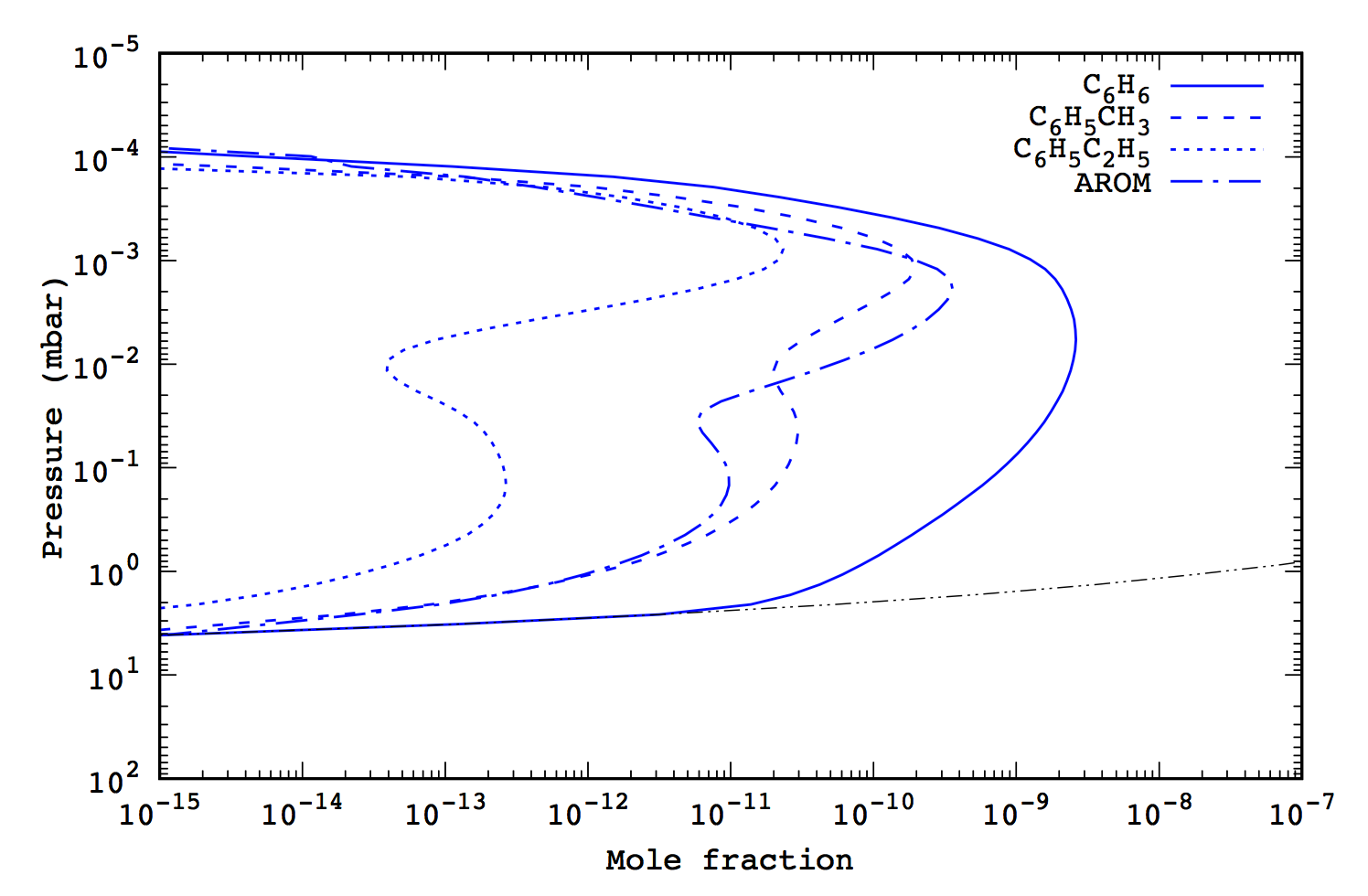} 
\caption{Mole fraction profiles of the major aromatics as a function of pressure derived from the coupled ion-neutral model (model A2, in blue). The dashed-dot curve in black corresponds to the saturation profile of benzene. The production of aromatics by the neutral model (A1) is very low and their profiles are not depicted.}
\label{fig:arom}
\end{center}
\end{figure}

In a neutral photochemical model, the production rate of aromatics and benzene in particular are quite low. The mole fraction of benzene is about $10^{-12}$ in \citet{Moses2005} at $10^{-3}$ mbar and about $10^{-13}$ in our present model (model A1) at the same pressure level. The introduction of ionic chemistry in the model has a strong effect on the production of aromatics in the atmosphere of Neptune, similar to the atmosphere of Titan \citep{Vuitton2018, Loison2018} (see Section \ref{section:aromatic_chemistry} for a dedicated discussion on the chemistry of aromatics). Figure \ref{fig:arom} shows the mole fraction profiles of the main aromatics in the model: benzene (C$_6$H$_6$), toluene (C$_6$H$_5$CH$_3$), ethylbenzene (C$_6$H$_5$C$_2$H$_5$) and the generic aromatic AROM (which is the sum of all other aromatic species not described in the present chemical scheme, see \citet{Loison2018}). Consequently, the detection of benzene and the determination of its abundance profile could give valuable constraints on the ionic chemistry in Neptune's atmosphere. The other aromatics would probably be difficult to detect from Earth-based observatories but could be detected by a putative in situ probe \citep{Mousis2018}. Only benzene has a high enough relative abundance at the saturation level to contribute significantly to haze production.

\begin{figure}[htp]
\begin{center}
\includegraphics[width=1.0\columnwidth]{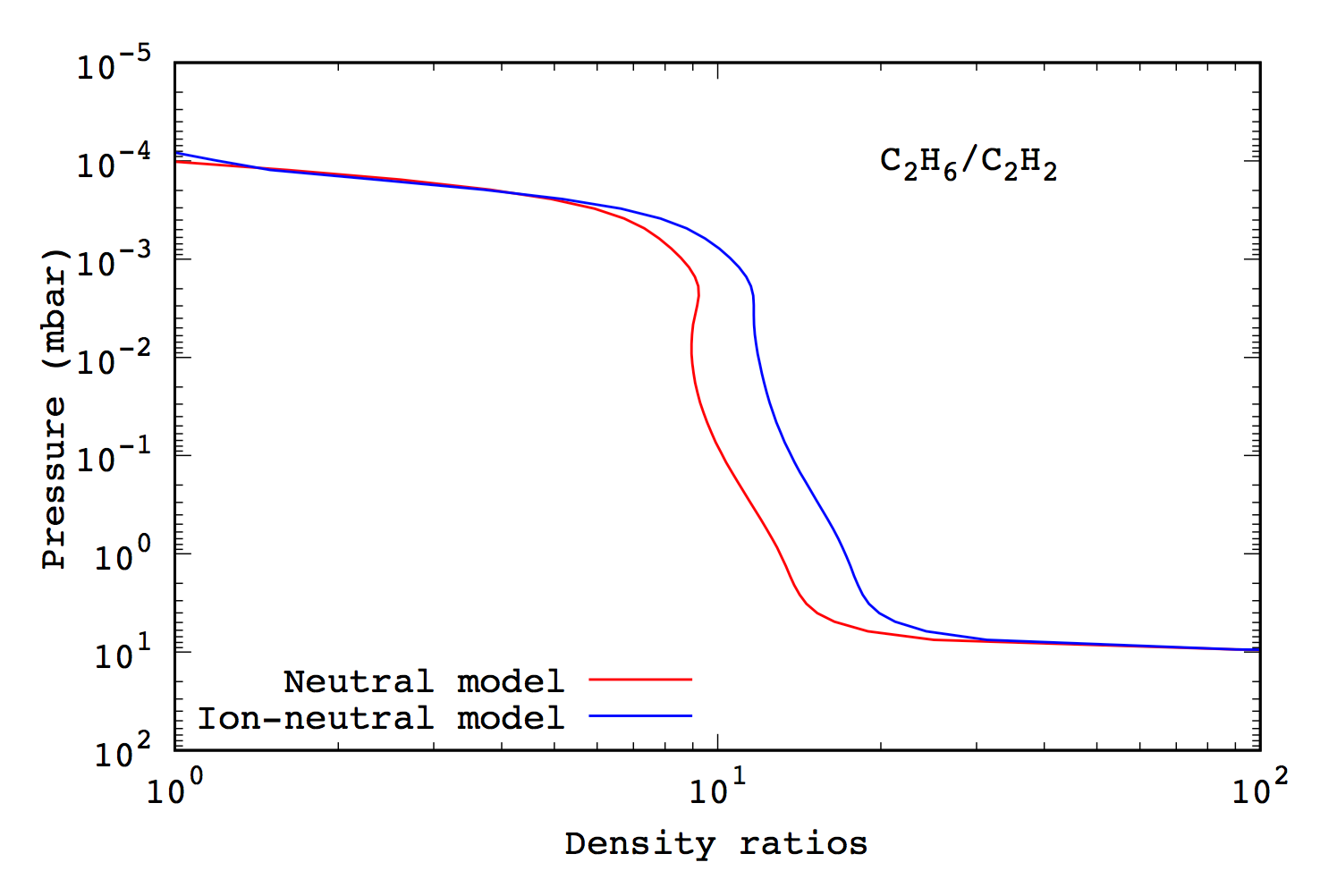} 
\caption{C$_2$H$_6$/C$_2$H$_2$ density ratios for the neutral model (model A1, in red) and the ion-neutral model (model A2, in blue) obtained from Figure \ref{fig:C2}. The increase of this ratio from model A1 to model A2 is mainly due to the decrease of the C$_2$H$_2$ abundance.}
\label{fig:C2H2_C2H6}
\end{center}
\end{figure}

The impact of ionic chemistry on the abundance of the main hydrocarbons is less pronounced. One noticeable point is that C$_2$H$_2$ is more affected by ionic chemistry than C$_2$H$_6$ (Figure \ref{fig:C2}). In Figure \ref{fig:C2H2_C2H6}, we show the C$_2$H$_6$/C$_2$H$_2$ density ratios for the neutral model and the ion-neutral model. Details of the chemistry of C$_2$H$_2$ are given in Section \ref{section:hydrocarbons_chemistry}.
From these results we can conclude that C$_2$H$_2$ is significantly depleted in a model that includes ionic chemistry, which may validate the hypothesis made by \citet{Hue2018} that ionic chemistry could cause the observed depletion seen in C$_2$H$_2$ (but not in C$_2$H$_6$) in the polar regions of Jupiter. Some other species are relatively abundant and affected by ionic chemistry like C$_4$H$_8$ and C$_6$H$_4$ (results not shown).

\subsection{Neutral oxygen compounds abundances}

Our neutral model (model A1) of oxygen species gives results very similar to the ones of \citet{Moses2018}. In the following, we only highlight our results obtained with the coupled ion-neutral models.

\subsubsection{Impact of ionic chemistry on oxygen compounds}

\begin{figure}[htp]
\begin{center}
\includegraphics[width=1.0\columnwidth]{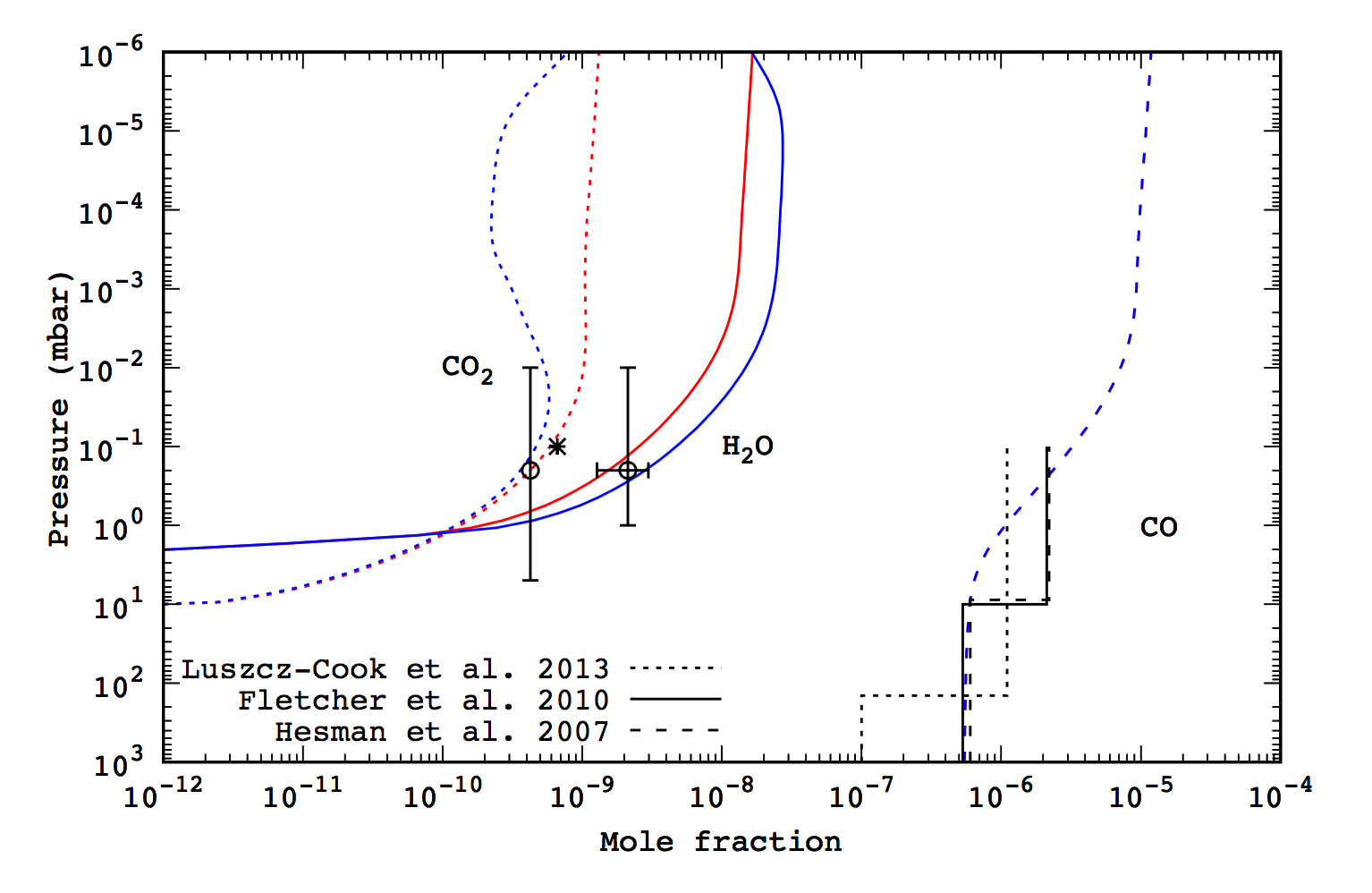} 
\caption{Mole fraction profiles for the neutral model (model A1, in red) and the ion-neutral model (model A2, in blue) of H$_2$O, CO and CO$_2$ considering an external origin for CO, CO$_2$ and H$_2$O as well as an internal origin of CO. For CO, the two model profiles overlap. Observations are given in black.  Star:  CO$_2$ \citep{Meadows2008}; Empty circles: CO$_2$ and H$_2$O \citep{Feuchtgruber1997}; Solid, dashed and dotted lines for CO \citep{Fletcher2010, Hesman2007, Luszcz-Cook2013} respectively.}
\label{fig:oxy}
\end{center}
\end{figure}

We see in Figure \ref{fig:oxy} that ionic chemistry increases the relative abundance of H$_2$O and decreases the one of CO$_2$. It has no effect on CO when assuming a constant flux of CO at the top of the atmosphere (model A2). Unfortunately, the main differences lie in regions (above $10^{-3}$ mbar pressure level) that are difficult to observe by remote instruments. On the other hand, determination of the abundance profile of CO$_2$ (from limb sounding with an orbiter, or in situ by a mass spectrometer) could give valuable information on the importance of ionic chemistry on oxygen species.

\subsubsection{The cometary impact hypothesis}

\begin{figure}[htp]
\begin{center}
\includegraphics[width=1.0\columnwidth]{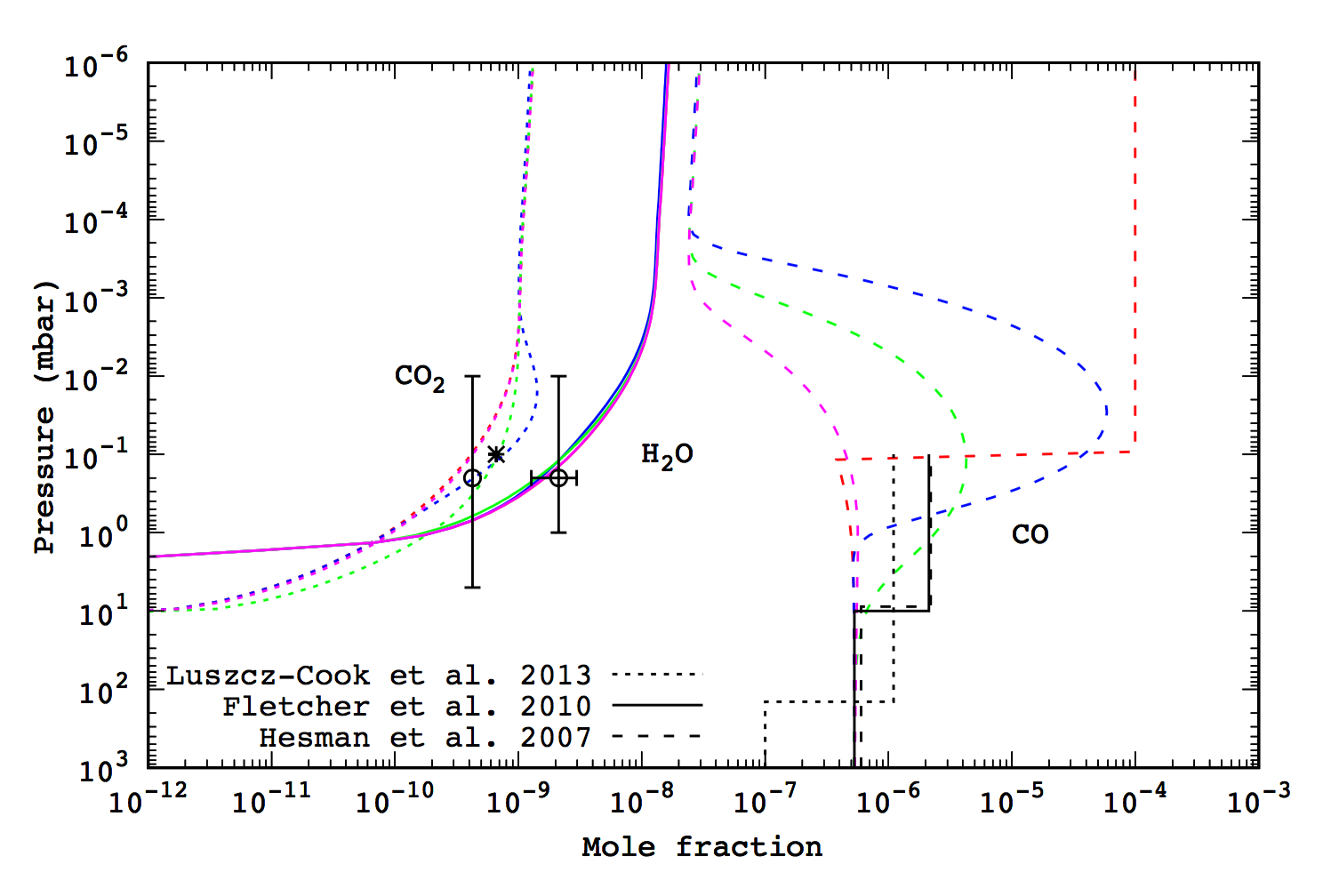} 
\caption{Time evolution of H$_2$O, CO and CO$_2$ mole fraction profiles, considering a cometary impact for the origin of CO in the upper atmosphere in addition to an internal origin of CO (model B2). In red: initial profile of CO due to the impact and the steady state of model B1. In blue, green and pink, the mole fractions of the three species at 5 years, 50 years and 500 years after the impact respectively. Observations are given in black (see previous caption for references).}
\label{fig:comet}
\end{center}
\end{figure}

In this section, we present the time evolution of CO, CO$_2$ and H$_2$O abundance profiles considering a cometary impact. Results are  presented in Figure \ref{fig:comet}. The mole fraction profiles of CO and CO$_2$ obtained at 5 and 50 years after the impact are roughly in agreement with observations, so we can expect that the various profiles obtained between these two dates could explain the actual abundances of CO and CO$_2$ (a careful study of these profiles using radiative transfer simulations is mandatory to go further into the discussion). Our model predicts that the CO$_2$ abundance increases after the cometary impact and then slightly decreases as a function of the decreasing CO abundance. After 50 years, both CO and CO$_2$ profiles disagree with observations. Recently, \citet{Teanby2019} have used Herschel/SPIRE observations to obtain abundances from multiple CO spectral features. They have obtained a CO profile in agreement with previous results with a nominal abundance in the troposphere of 0.22 ppm and 1 ppm in the stratosphere. 
As a consequence, our model favors a quite recent cometary impact (less than 50 years). We also see that the water profile cannot be used to decipher the origin of CO. 
This result seems to be quite different from the results of \citet{Moses2017}, which favor a cometary impact 200 years ago for the same impactor (see section \ref{section:UP} for further discussion). 

\subsection{Main ion densities}

\begin{figure}[htp]
\begin{center}
\includegraphics[width=1.0\columnwidth]{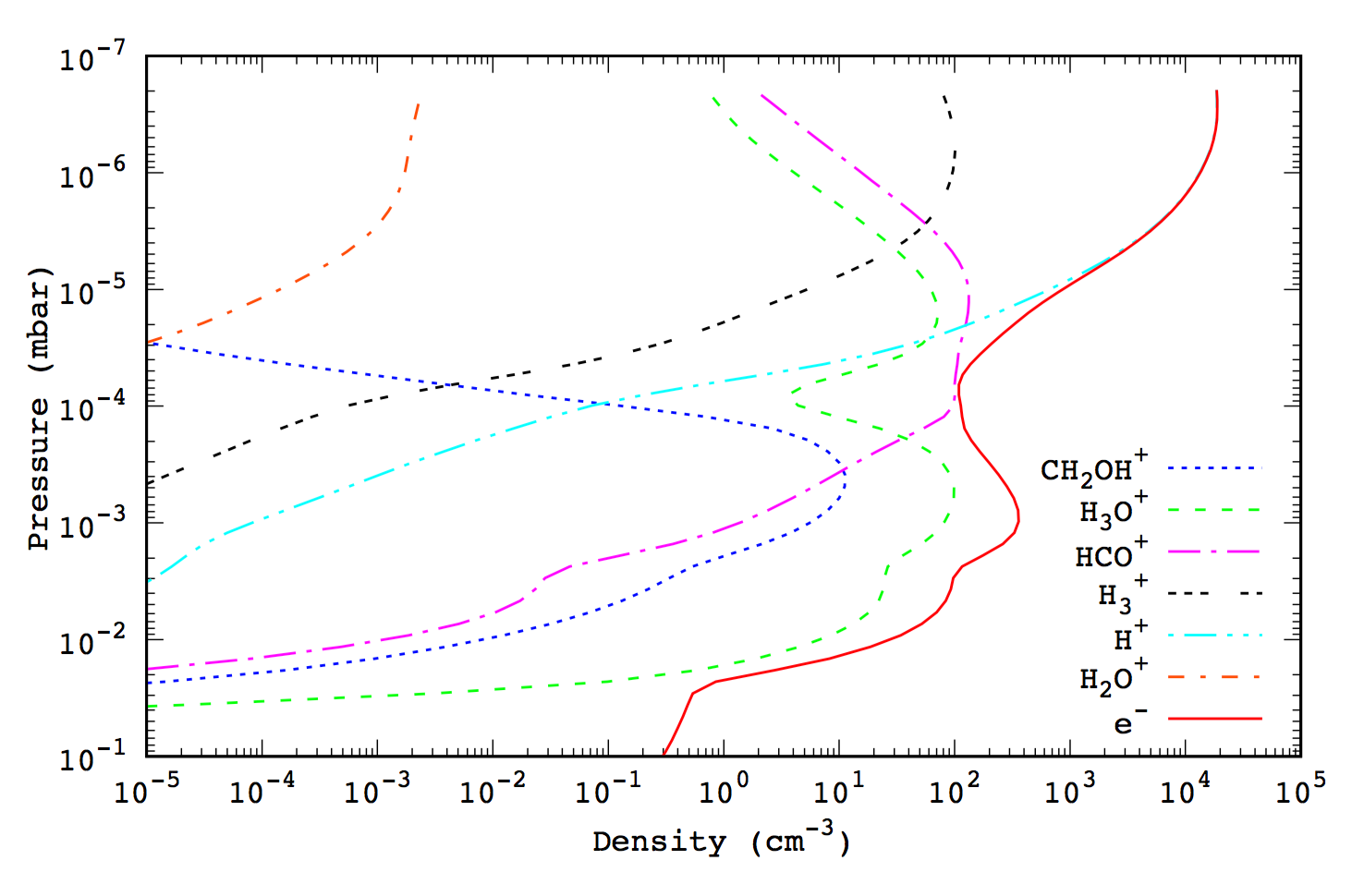} 
\includegraphics[width=1.0\columnwidth]{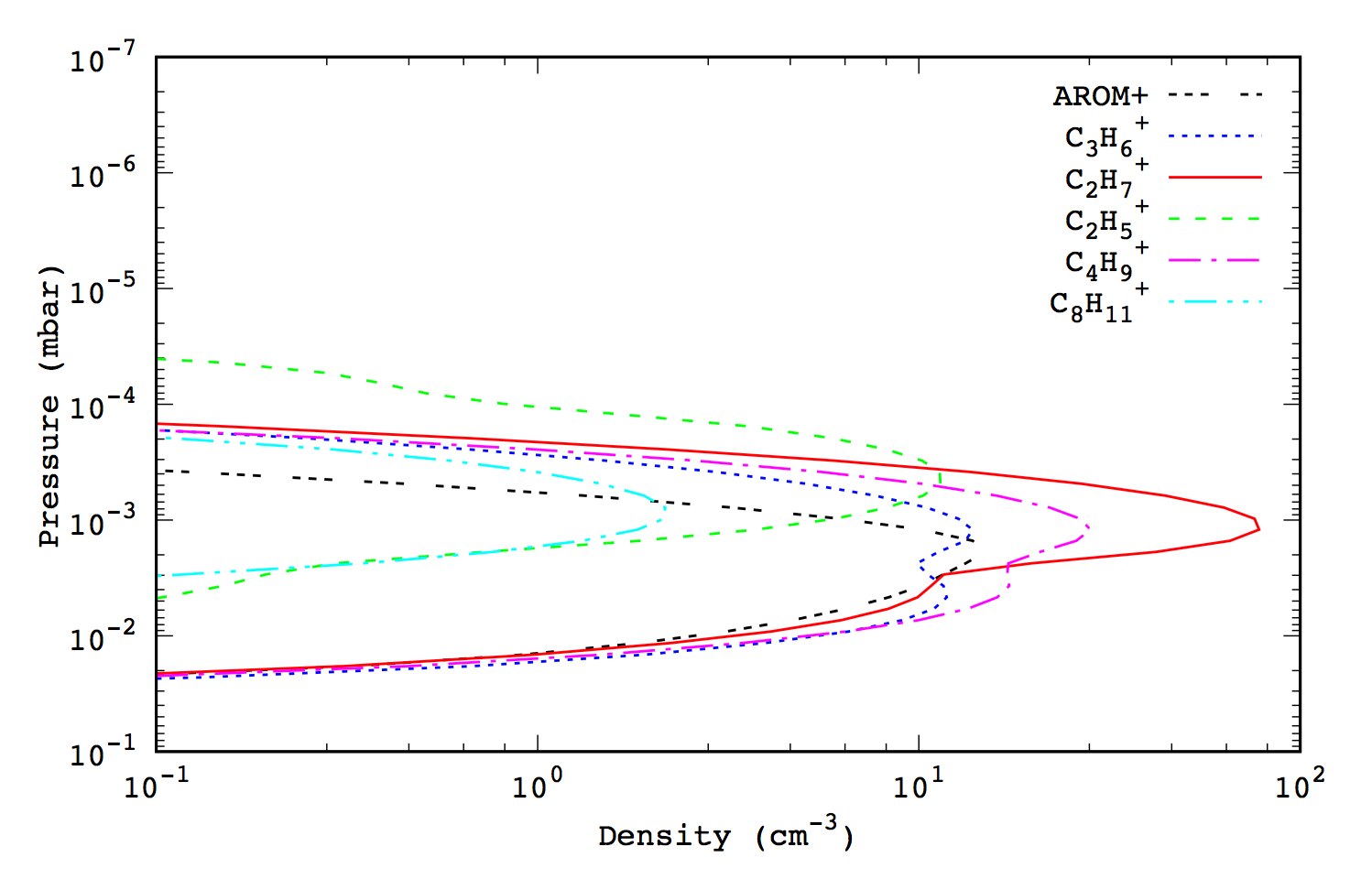}
\caption{Density profiles of the major ions as a function of pressure levels derived from the model A2.}
\label{fig:ions1}
\end{center}
\end{figure}

\begin{figure}[htp]
\begin{center}
\includegraphics[width=1.0\columnwidth]{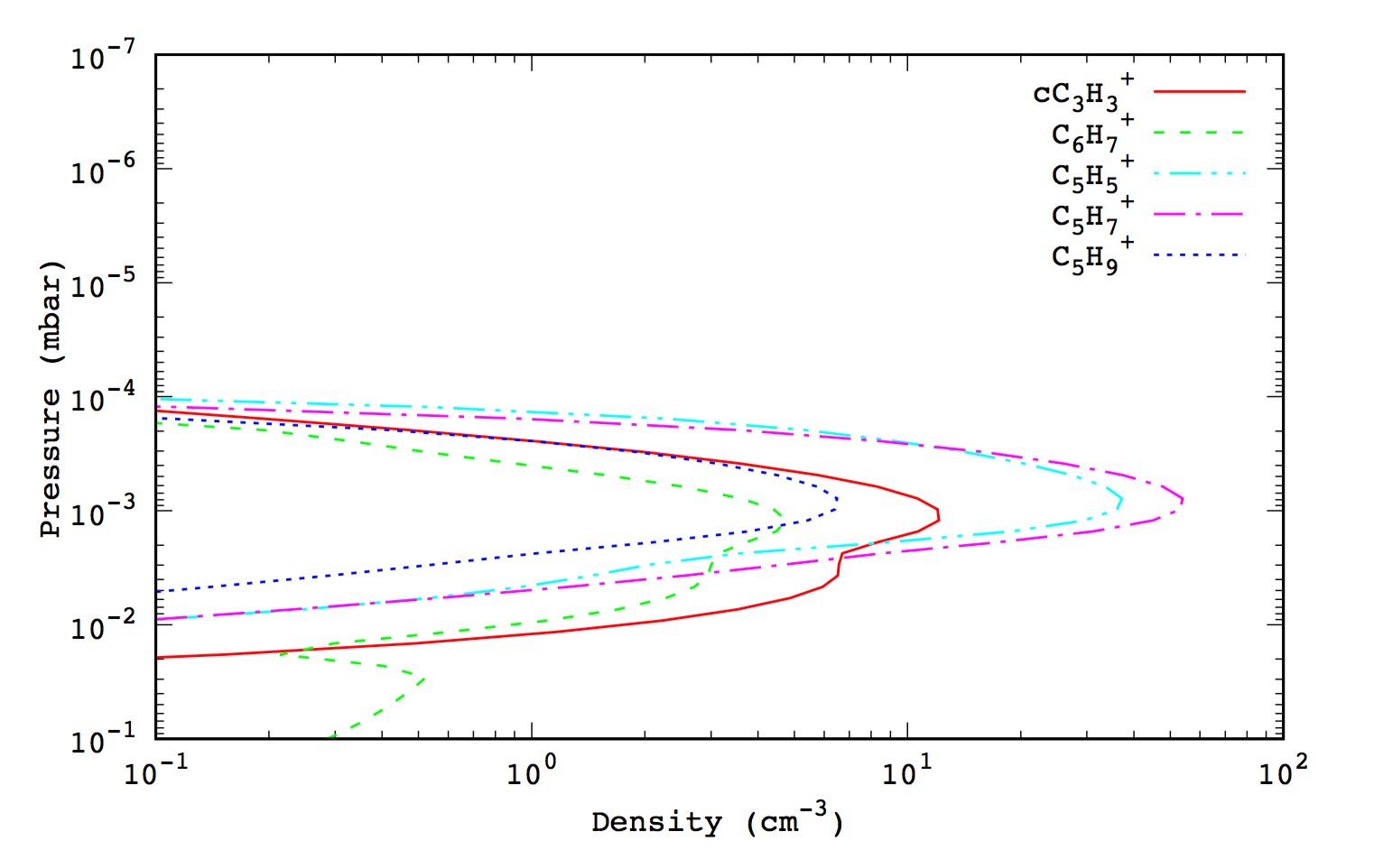} 
\caption{Density profiles of the major ions as a function of pressure levels derived from the model A2.}
\label{fig:ions2}
\end{center}
\end{figure}

\begin{figure}[htp]
\begin{center}
\includegraphics[width=1.0\columnwidth]{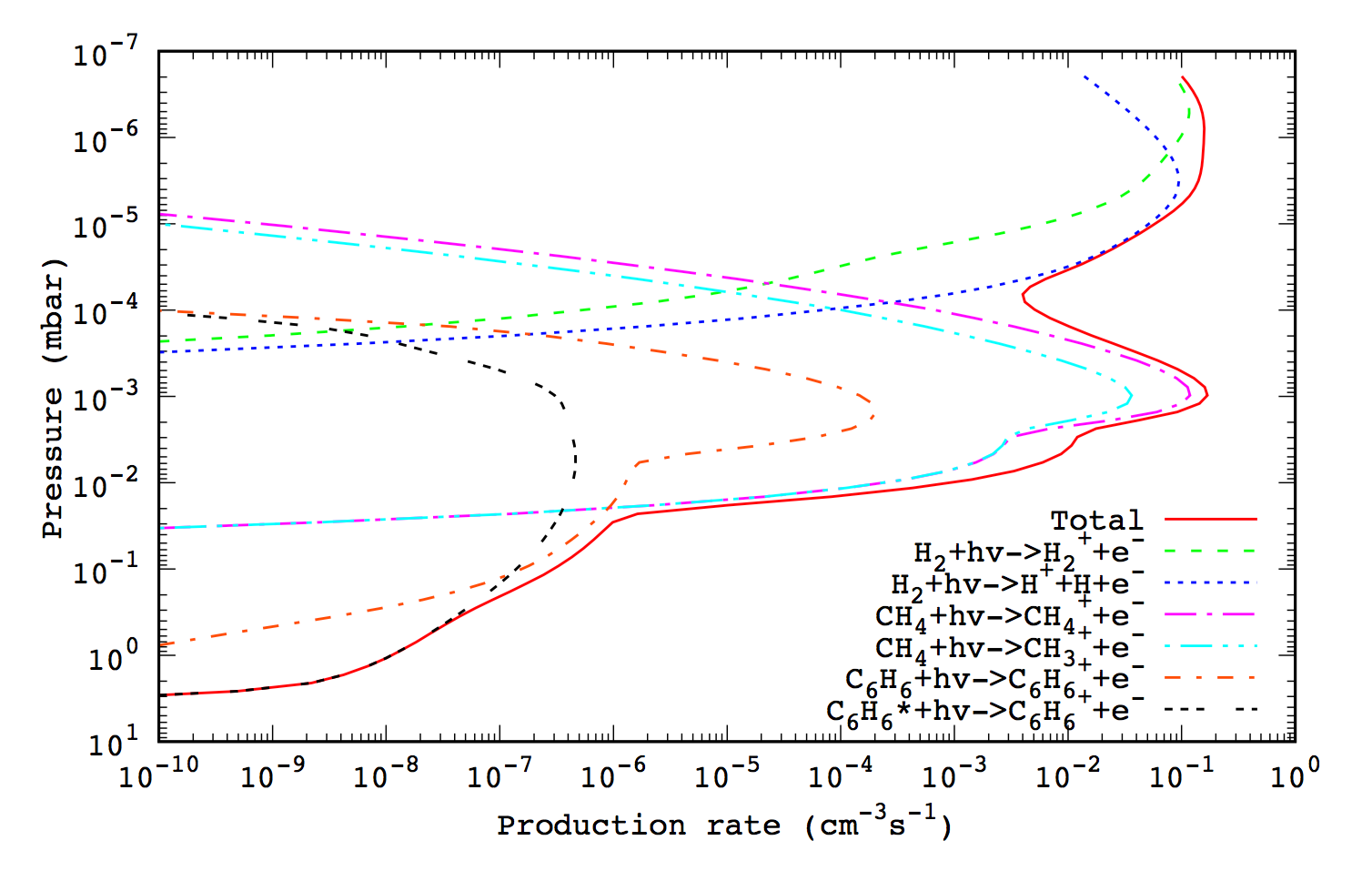} 
\includegraphics[width=1.0\columnwidth]{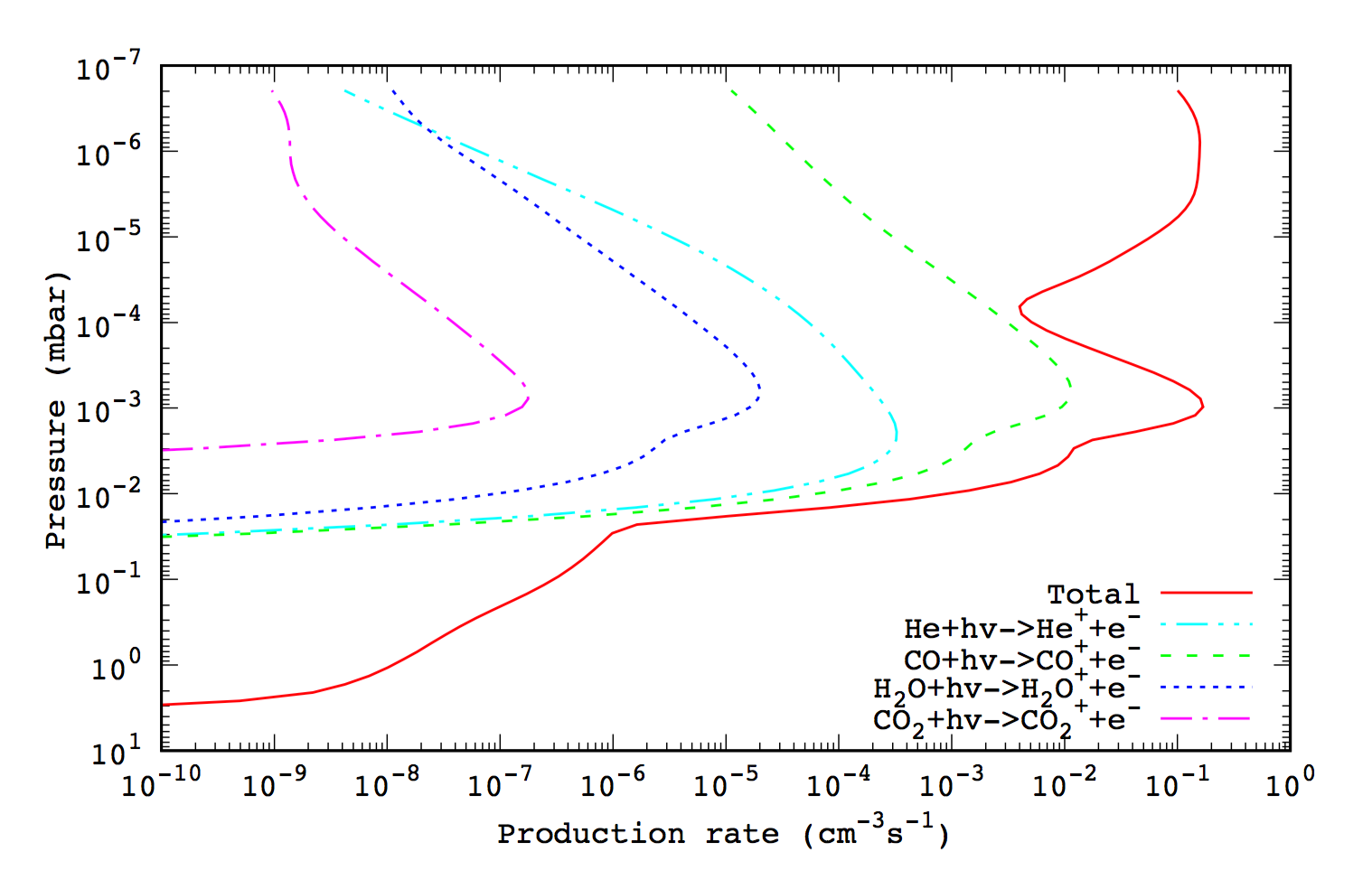}
\caption{Total production rate of electrons (in red) and relative production rate of the main reactions of electron production as a function of pressure in model A2. Top: Relative production rate of electrons from ionisation of H$_2$, CH$_4$ and C$_6$H$_6$. Bottom: Relative production rate of electrons from ionisation of He, CO, H$_2$O and CO$_2$.}
\label{fig:rates}
\end{center}
\end{figure}

Our ion-neutral model gives the density profiles of the main hydrocarbon ions in the stratosphere of Neptune (see Figures \ref{fig:ions1} and \ref{fig:ions2}). We identify two main ion layers: 
(i) in the upper atmosphere (above the $10^{-4}$ mbar pressure level), H$^+$, H$_3^+$, HCO$^+$ and H$_3$O$^+$ are the main ions. H$^+$ is the most abundant ion above $10^{-5}$ mbar with a density of about $2\times 10^4$ cm$^{-3}$, equal to the electron concentration. This result is slightly greater than the measurements inferred from data acquired during the Voyager 2 RSS occultation \citep{Lindal1992} with an electron concentration of $7\times 10^3$ cm$^{-3}$ at 1200 km (near a latitude of 45$^\circ$S), corresponding to a pressure level of $3.3\times 10^{-7}$ mbar in our model. (ii) Around $10^{-3}$ mbar, the ion layer is composed of H$_3$O$^+$, hydrocarbons (mainly C$_2$H$_7^+$) and aromatics. The production of many ions and electrons (see Figures \ref{fig:ions1} and \ref{fig:ions2}) is relatively efficient down to pressure as low as 0.01 mbar. Below this level, the production of electrons is dominated by the photoionization of C$_6$H$_6$ (see Figure \ref{fig:rates}).

The two main regions of electron production are illustrated in Figure \ref{fig:rates}. They are dominated by the photolysis of H$_2$, CH$_4$ and C$_6$H$_6$. For comparison, the production rate of electrons from the photoionisation of CO, H$_2$O and CO$_2$ are also presented (in case of model A2 with a high abundance of CO in the upper atmosphere). Their maximum occurs at the same level (around $10^{-3}$ mbar). However, we see that, despite the high abundance of CO in Neptune's atmosphere, these photoionisation processes do not contribute significantly to the ionic chemistry.

\citet{Lyons1995} developed a 1D-photochemical model, based on the model of \citet{Romani1993}, that accounted for eddy and molecular diffusion, and included ion and neutral chemistry of hydrocarbons, water and related species, and "metals" (species that do not contain C, H, O or N) produced by meteoroid ablation. The peak locations of electrons in the model of \citet{Lyons1995} are roughly in agreement with our model results with a peak at $10^{-6}$ mbar and another one around $10^{-3}$ mbar. However, the peak densities are quite different (2 times lower at $10^{-6}$ mbar and 3-4 times higher at $10^{-3}$ mbar in their model). Also, the density profile of H$_3$O$^+$ is quite different from ours and we find that HCO$^+$ is the major ion around $10^{-4}$ mbar, which was not present in the \citet{Lyons1995} model. 
In their model, the mole fraction of CH$_4$ at the lower boundary was 10 times lower than the value accepted now and the eddy diffusion coefficient used was very different than the one used in the present study (about $10^3$ times higher than the value we use at 0.1 mbar). Also, the chemical scheme has been strongly improved since the model of \citet{Romani1993}. Consequently, our results for hydrocarbons and oxygen-bearing species are very different. \citet{Lyons1995} predicted that singly ionized magnesium was the most likely metal to be found in the layer around $10^{-3}$ mbar pressure level. In our model, the main ions are H$_3$O$^+$ and C$_2$H$_7^+$ (and other heavier hydrocarbons) with a total density 3-4 times lower at that pressure level. Considering the strong difference between the model of \citet{Lyons1995} and ours for most of the neutral and ion species, it might be interesting to update the model of metals he developed to confirm if Mg$^+$ is the dominant species around $10^{-3}$ mbar.
According to the radio occultation data from Voyager 2 analyzed by \citet{Lindal1992}, our model underestimates the production of electrons in the stratosphere. This might be an indication that magnetospheric electrons and galactic cosmic rays should be included in the model. However, as stated by \citet{Lindal1992}, the densities calculated in the lower ionosphere are quite uncertain. Confirmation of these results would be of great importance to better constrain the models.

\subsection{Simulated mass spectrometer spectra}

\begin{figure}[htp]
\begin{center}
\includegraphics[width=0.8\columnwidth]{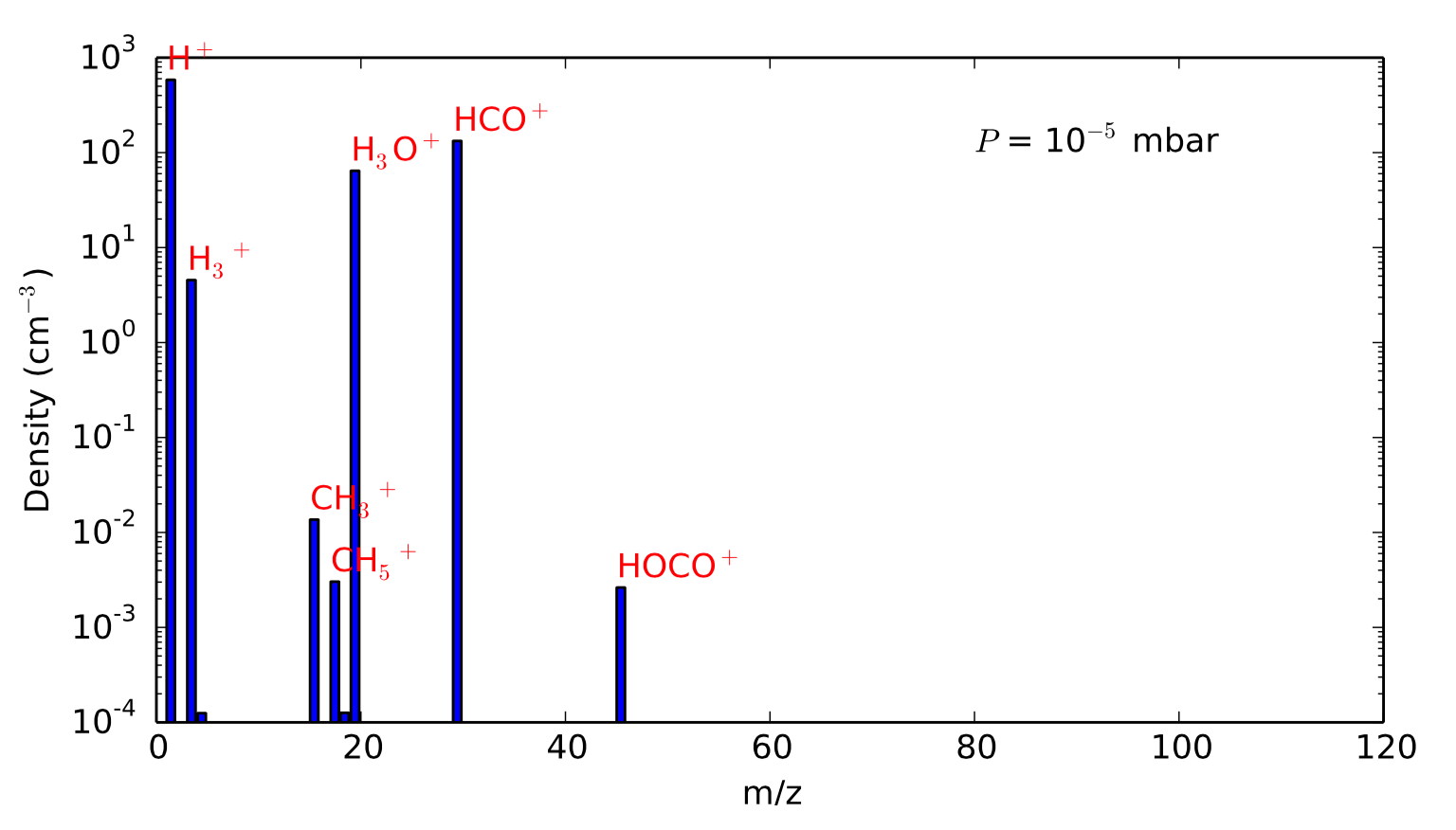} 
\includegraphics[width=0.8\columnwidth]{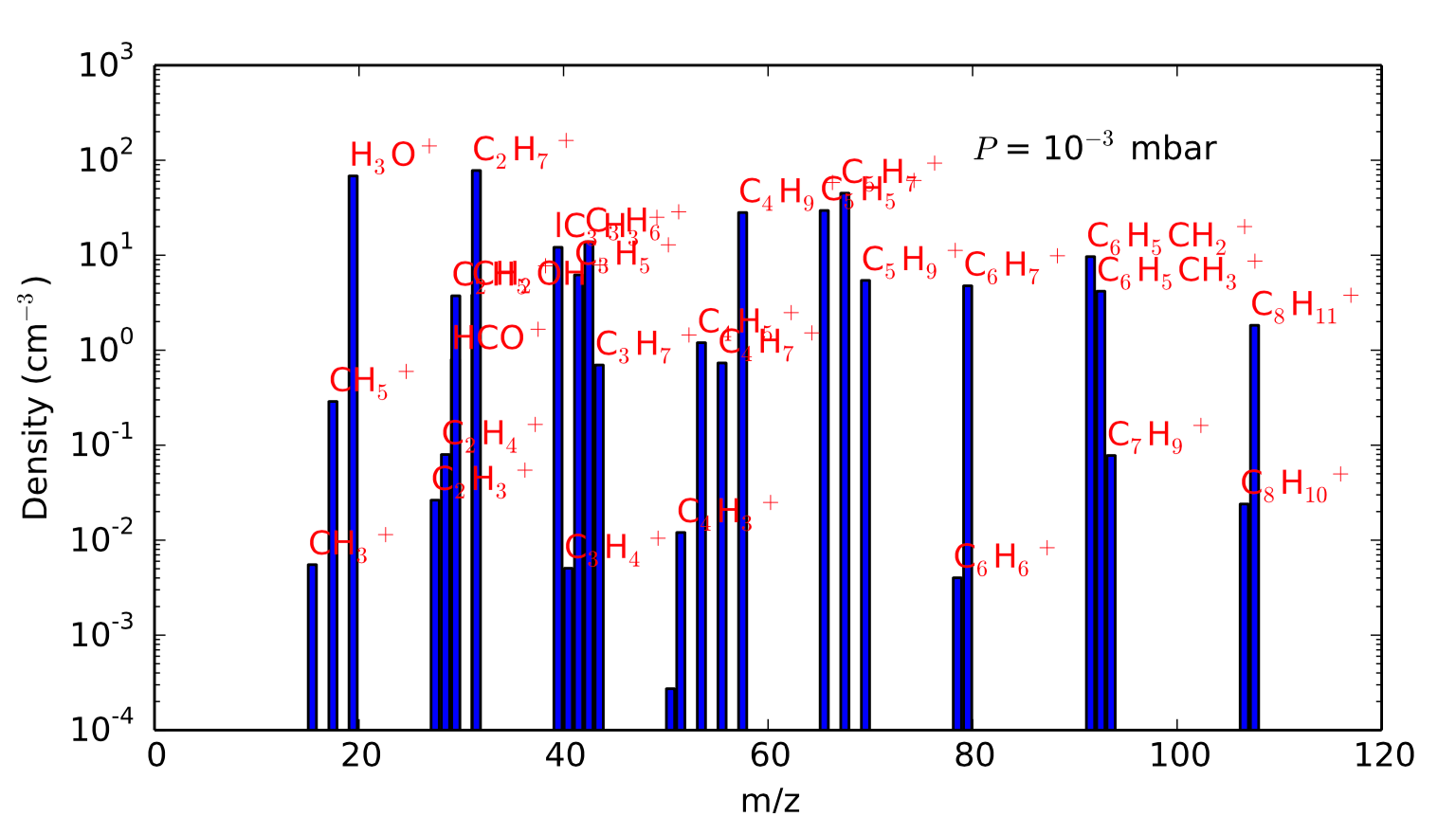}
\includegraphics[width=0.8\columnwidth]{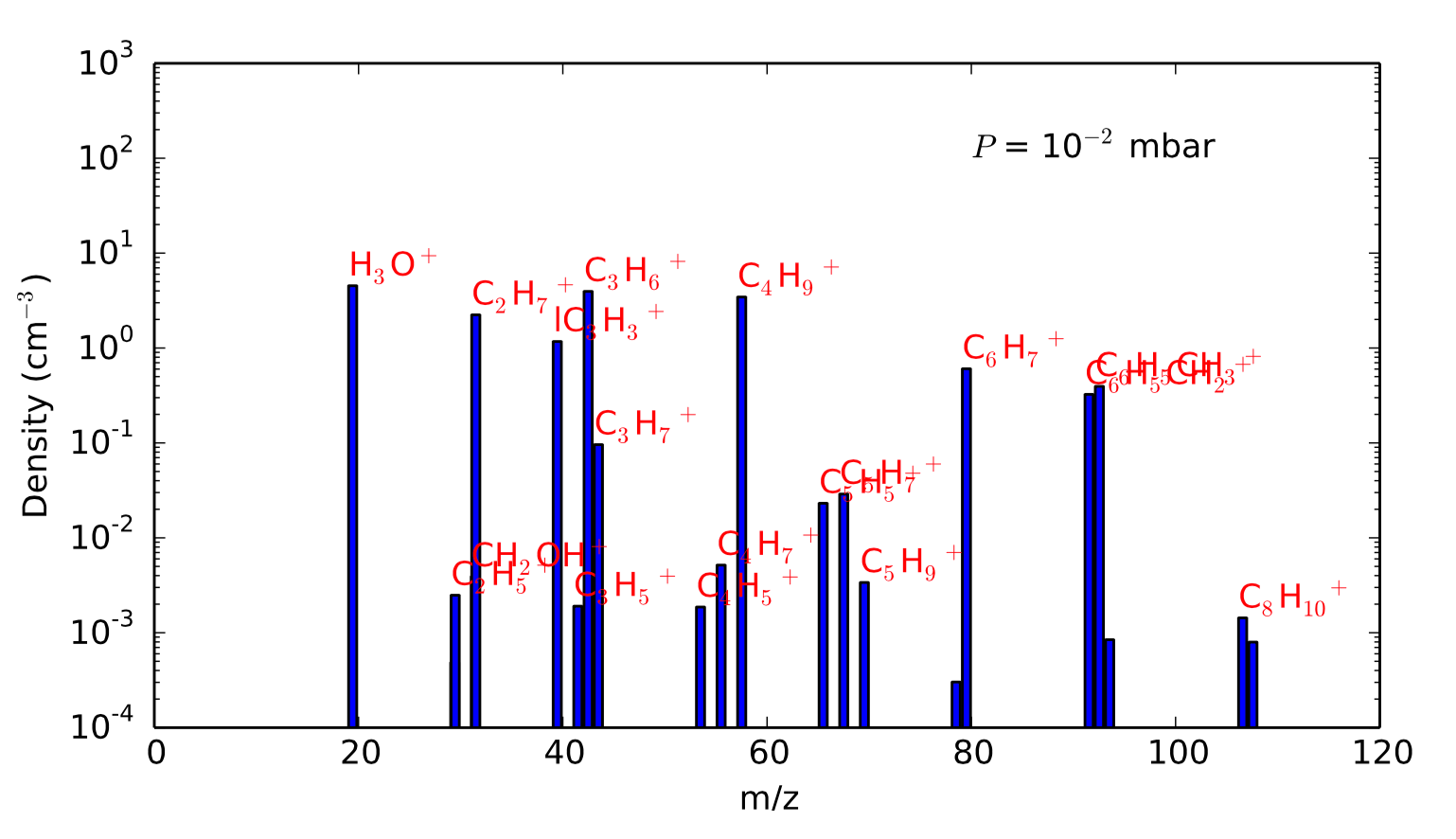}
\caption{Simulated mass spectra obtained from our model results (model A2) at three different pressure levels.}
\label{fig:mass}
\end{center}
\end{figure}

Recently, \citet{Mousis2018} discussed the importance of a dedicated mission to Uranus and/or Neptune exploration and in particular the need to send an atmospheric probe to determine, among many other physical and chemical properties, the stratospheric temperature and the composition in hydrocarbons and other chemical species. Such data would offer unprecedented constraints for photochemical models. To maximise the science return, it will be necessary to build a mass spectrometer with a high mass resolution allowing all the species to be distinguished. One major contribution of such an instrument would be to measure the abundances of noble gases and key isotope ratios, and sample atmospheric regions far below those accessible to remote sensing. In the present study, we highlight the importance of such a probe to determine the abundance profile of many species in the ionosphere and the stratosphere. In Figure \ref{fig:mass}, we show three simulated mass spectra for ions only obtained from our model results (model A2) with an ideal mass resolution (limited by the uncertainty on the atomic mass of C, H and O) at three different pressure levels. Ions obtained from the fragmentation of neutrals are not considered in this case. To our knowledge, there is no probe concept published so far that could obtain such mass spectra at the expected pressure levels between 0.1 and $10^{-6}$ mbar of a giant planet.

We see that many species could be detectable, provided that mass resolution and sensitivity are in agreement with our simulated spectra. While light species like H$_2^+$ and HCO$^+$ would bring crucial information regarding the processes occurring in the ionosphere and the origin of CO, heavy species in the stratosphere would provide clues on the complexity of hydrocarbon photochemistry. In particular, our model predicts that aromatic ions are relatively abundant in the middle stratosphere. We note also that H$_3$O$^+$ might be detectable throughout the stratosphere. These kind of data, combined with neutral mass spectra, would offer the possibility to detect many species (from C$_2$- up to C$_8$- compounds in our model) and to determine their abundances; Data which are clearly lacking right now to better constrain photochemical models.

\subsection{Chemistry overview}

\subsubsection{Main hydrocarbons chemistry}
\label{section:hydrocarbons_chemistry}

\begin{figure}[htp]
\begin{center}
\includegraphics[width=1.0\columnwidth]{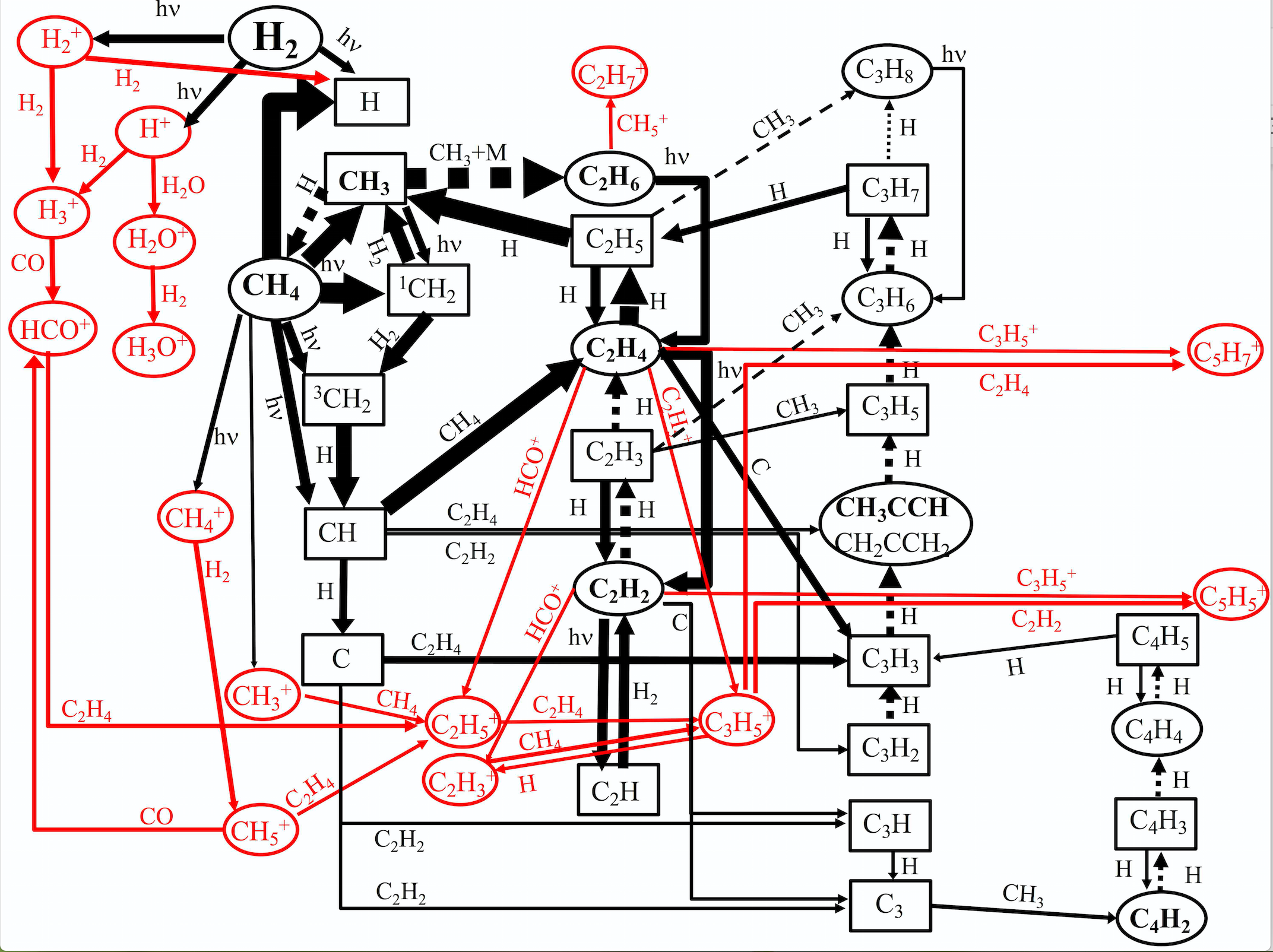} 
\caption{Schematic diagrams highlighting the important neutral (black) and ionic reaction (red) pathways for the atmospheric chemistry of Neptune. The thickness of each arrow is proportional to the integral of the total production rate over the atmosphere. Radicals are shown in boxes, whereas stable compounds and ions are shown in circles. Dashed lines correspond to three body reactions. Species in bold correspond to the ones that have been detected so far.}
\label{fig:scheme2}
\end{center}
\end{figure}

In order to summarize the main chemical pathways in our model, we depict in Figure \ref{fig:scheme2} a schematic diagram based on the integral over the entire altitude range of the production rate of each reaction given in Appendix \ref{appendix:rates}. These rates correspond to the steady state of model A2. There are many similarities with the photochemistry of Titan except for the ionic chemistry of H$_3^+$ (and related species). Also, C$_4$H$_2$ is mainly produced by the reaction C$_3$ + CH$_3$ (and not by C$_2$H + C$_2$H$_2$ as is the case for Titan's atmosphere). It should be noted that the reactivity of C$_5$H$_5^+$ isomers is not well known (Ozturk 1989) and their abundance may be overestimated in our model. H$_3$O$^+$, C$_2$H$_7^+$, C$_5$H$_5^+$ and C$_5$H$_7^+$ are quite abundant in our model because they react mainly with electrons (not represented in the diagram). 

\begin{figure}[htp]
\begin{center}
\includegraphics[width=0.5\columnwidth]{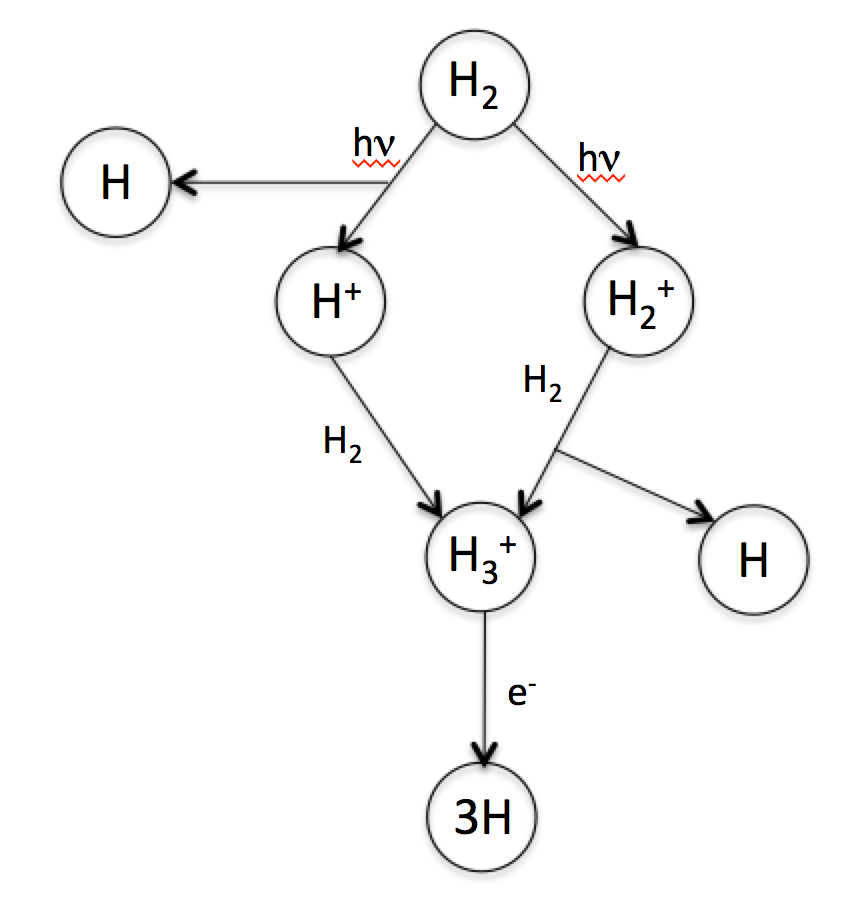} 
\caption{Photochemical scheme of atomic hydrogen production in the ionosphere of Neptune.}
\label{fig:scheme1}
\end{center}
\end{figure}

The chemistry of C$_2$H$_2$ in Neptune's atmosphere is somewhat different than the one in Titan's atmosphere. In particular, the abundance of HCN is lower in Neptune's atmosphere than in Titan's.  Then, instead of reacting with HCN (leading to C$_2$H$_4$ + HCNH$^+$),  C$_2$H$_5^+$ reacts mainly with C$_2$H$_4$ leading to C$_3$H$_5^+$, which does not give back C$_2$H$_4$. This has an influence on C$_2$H$_2$ abundances as C$_2$H$_2$ results mainly from the photodissociation of C$_2$H$_4$. In addition, to reduce the production of C$_2$H$_4$, ionic chemistry increases the loss of C$_2$H$_2$ and C$_2$H$_4$ through protonation reactions by HCO$^+$ (reactions involving C$_2$H$_3^+$ and C$_2$H$_5^+$ give only a small amount of C$_2$H$_2$ and C$_2$H$_4$). In addition, the low abundance of HCN induces a higher abundance of C$_2$H$_5^+$ (due to the low efficiency of the C$_2$H$_5^+$ + HCN reaction), and then a higher consumption of C$_2$H$_4$ by the C$_2$H$_4$ + C$_2$H$_5^+$ reaction. Also, the decrease in the mole fraction of C$_2$H$_2$ (and C$_2$H$_4$) in the ion-neutral model (model A2) is due to the increase in the production of atomic hydrogen by ionic chemistry (following the scheme depicted in Figure \ref{fig:scheme1}), which then reacts with C$_2$H$_2$ (and C$_2$H$_4$) to increase its overall loss. The mole fraction profiles of H for models A1 and A2 are presented in Figure \ref{fig:CH4}.

\subsubsection{Aromatics chemistry}
\label{section:aromatic_chemistry}

\begin{figure}[htp]
\begin{center}
\includegraphics[width=1.0\columnwidth]{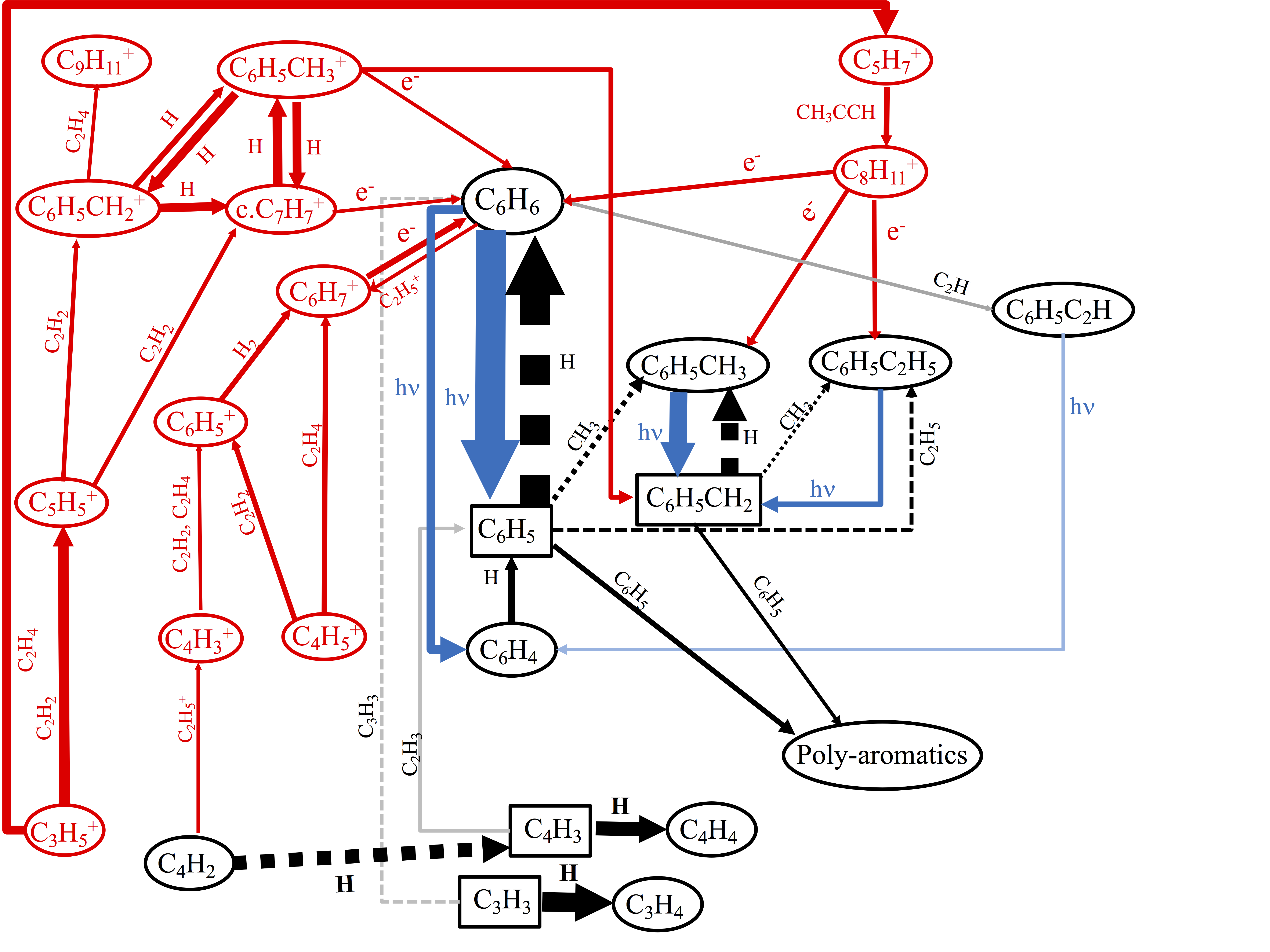} 
\caption{Schematic diagram highlighting the important neutral (in black) and ionic (in red) pathways for the production of aromatic species in Neptune's atmosphere. The photodissocation of C$_6$H$_6$  has been simplified for the clarity of the figure (see \citet{Loison2018} for details on C$_6$H$_6$ photodissociation and the role of metastable C$_6$H$_6$). Photodissociations are in blue. The thickness of each arrow is proportional to the integral of the total production rate over the atmosphere (see Appendix \ref{appendix:rates}). The pale grey and blue lines represent very low fluxes. Three body reactions are represented by dashed lines. Radicals are shown in boxes, whereas closed shell compounds are shown in circles.}
\label{fig:scheme3}
\end{center}
\end{figure}

Aromatics are produced mainly through ionic reactions, C$_3$H$_5^+$ and C$_4$H$_5^+$ being the main precursors of aromatics in Neptune's atmosphere as shown in the Figure \ref{fig:scheme3}. It should be noted that the neutral radical chemistry is very ineffective in Neptune's atmosphere because C$_3$H$_3$ and C$_4$H$_3$ radicals react mainly with atomic hydrogen and also because C$_4$H$_2$ has a much smaller role than in Titan's atmosphere. It should also be noted that, except for C$_6$H$_7^+$, the electronic dissociation recombinations producing aromatic compounds at work in the model are very poorly known, which generate large uncertainties. Moreover, there are other sources of uncertainties (already present in the model of Titan \citep{Loison2018}). First, the generic species AROM, representing all aromatics not explicitly described in the chemical scheme (such as C$_6$H$_5$C$_2$H$_3$ and C$_6$H$_5$-C$_6$H$_5$) reaches a relatively high abundance, which necessarily leads to bias in the results. Secondly, we have not introduced reactions between C$_6$H$_5$ and C$_4$H$_4$ and C$_4$H$_6$ that could be rapid, even at low temperatures, producing Polycyclic Aromatic Hydrocarbons (PAH) \citep{Kaiser2012, Mebel2017}). These reactions could consume C$_6$H$_5$ and then reduce the concentration of C$_6$H$_6$. However, the total amount of aromatics will stay almost constant.

\section{Discussion}

\subsection{Effect of a large CO abundance on hydrocarbons}

\begin{figure}[htp]
\begin{center}
\includegraphics[width=1.0\columnwidth]{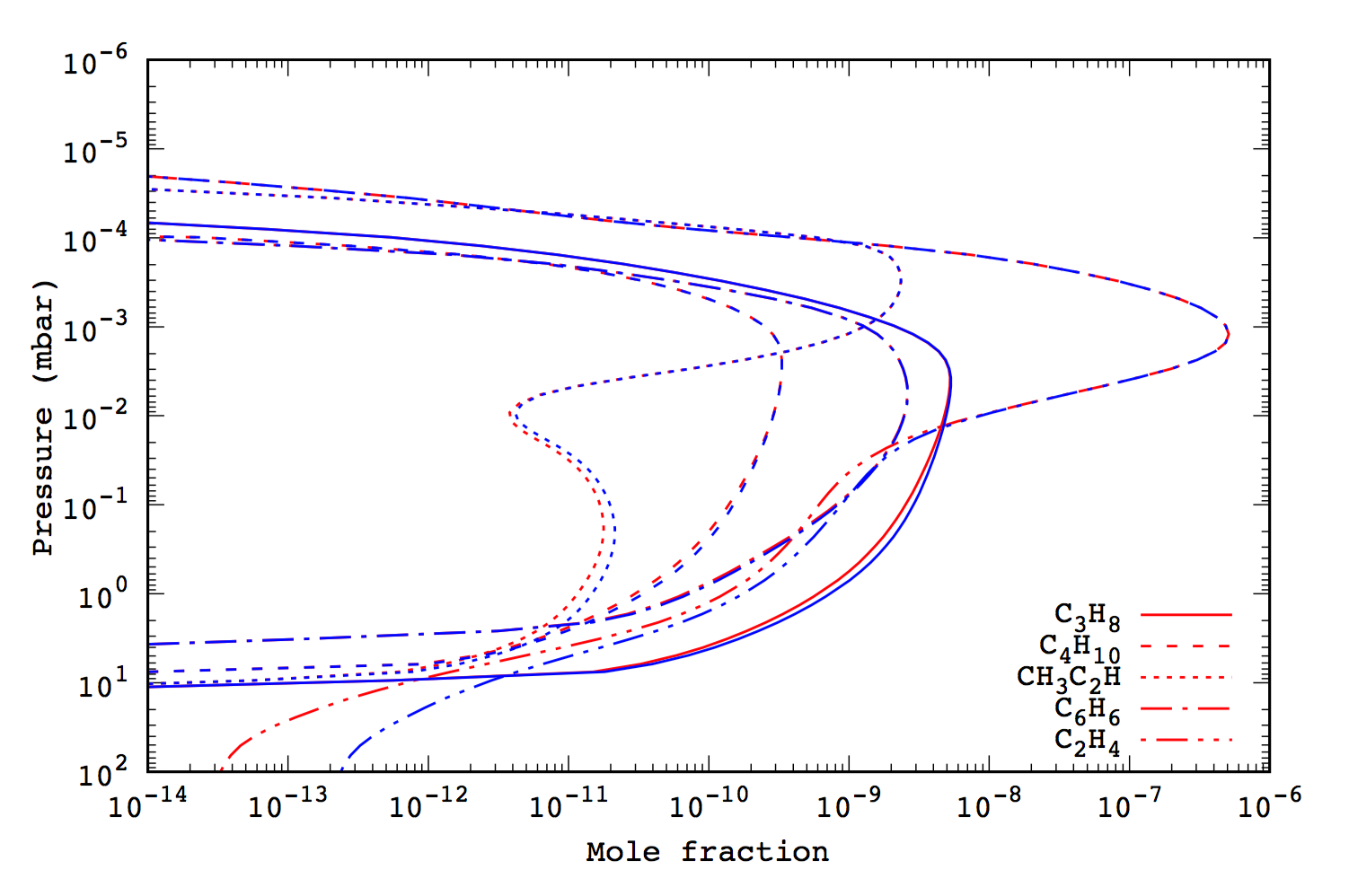} 
\caption{Mole fractions of several hydrocarbons obtained with the neutral model to test the influence of high CO abundance in the higher part of the atmosphere. In red: model A1. In blue: neutral model with no CO influx at the upper boundary.}
\label{fig:hi_CO}
\end{center}
\end{figure}

\citet{Moses2017} discussed the impact of a large CO mixing ratio in the stratosphere of Neptune on the mixing ratios of hydrocarbons in their neutral photochemical model. In particular, they found that a larger CO abundance leads to smaller mixing ratios of C$_2$H$_2$ and some higher-order hydrocarbons. Using our neutral model, we do not see any significant difference for C$_2$H$_x$ hydrocarbons and the impact on some other hydrocarbons is quite limited (see Figure \ref{fig:hi_CO}). \citet{Moses2017} argued that their results could be an artifact of the low-resolution ultraviolet cross sections used in their model. In the present model we have used high-resolution cross sections for most of the absorbing species (see \citet{Loison2017} for details on oxygen species). We have also tested low-resolution (1 nm) and high-resolution (0.1 nm) absorption cross-sections of CO on the entire range of absorption wavelengths without finding any significant difference for hydrocarbons. A further dedicated study seems necessary to clarify this point. Note that \citet{Kim2014} showed that the high-resolution H$_2$ cross sections have a strong influence on the results for Saturn's ionosphere. These results highlight the need for high-resolution cross sections for calculating the actinic flux in giant planet atmospheres. 

\begin{figure}[htp]
\begin{center}
\includegraphics[width=1.0\columnwidth]{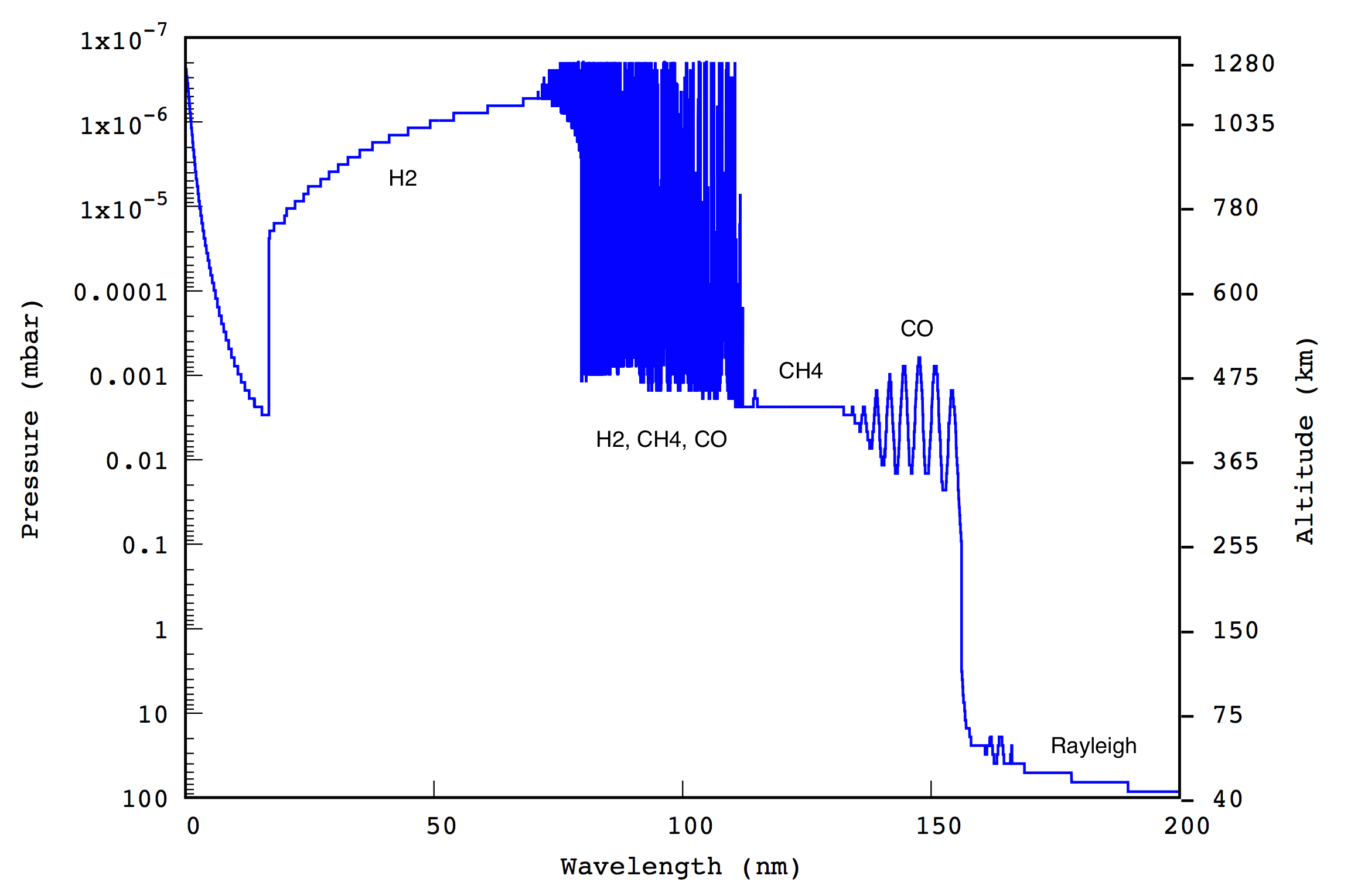} 
\caption{Depth of penetration of solar UV radiation in the atmosphere of Neptune as a function of pressure levels (model A2). The main species which are responsible for the total absorption are given.}
\label{fig:penetration}
\end{center}
\end{figure}

Figure \ref{fig:penetration} shows the depth of penetration of solar UV radiation as a function of altitude for the model A2. The curve represents the altitude at which the optical depth is equal to 1. The major species that are responsible for the total absorption as a function of wavelength are depicted. Between around 80 nm and 110 nm, the three species H$_2$, CH$_4$ and CO are the major absorbers, with all three having several discrete transitions throughout this range. We see also that CO is an efficient absorber around 150 nm, which is the wavelength region where C$_2$H$_x$ species dissociate. As a consequence, CO partially shields hydrocarbons from photolysis. Note that CO absorbs but is not photodissociated above 108 nm (CO fluoresces above the limit of dissociation). As a consequence, the abundance profile of CO may have a strong influence on the photochemistry around 400 km of altitude (0.01 mbar pressure level). Although it is limited for neutral species, it is quite pronounced for ions (see below). 

\subsection{Effect of a large CO abundance on ions}

\begin{figure}[htp]
\begin{center}
\includegraphics[width=1.0\columnwidth]{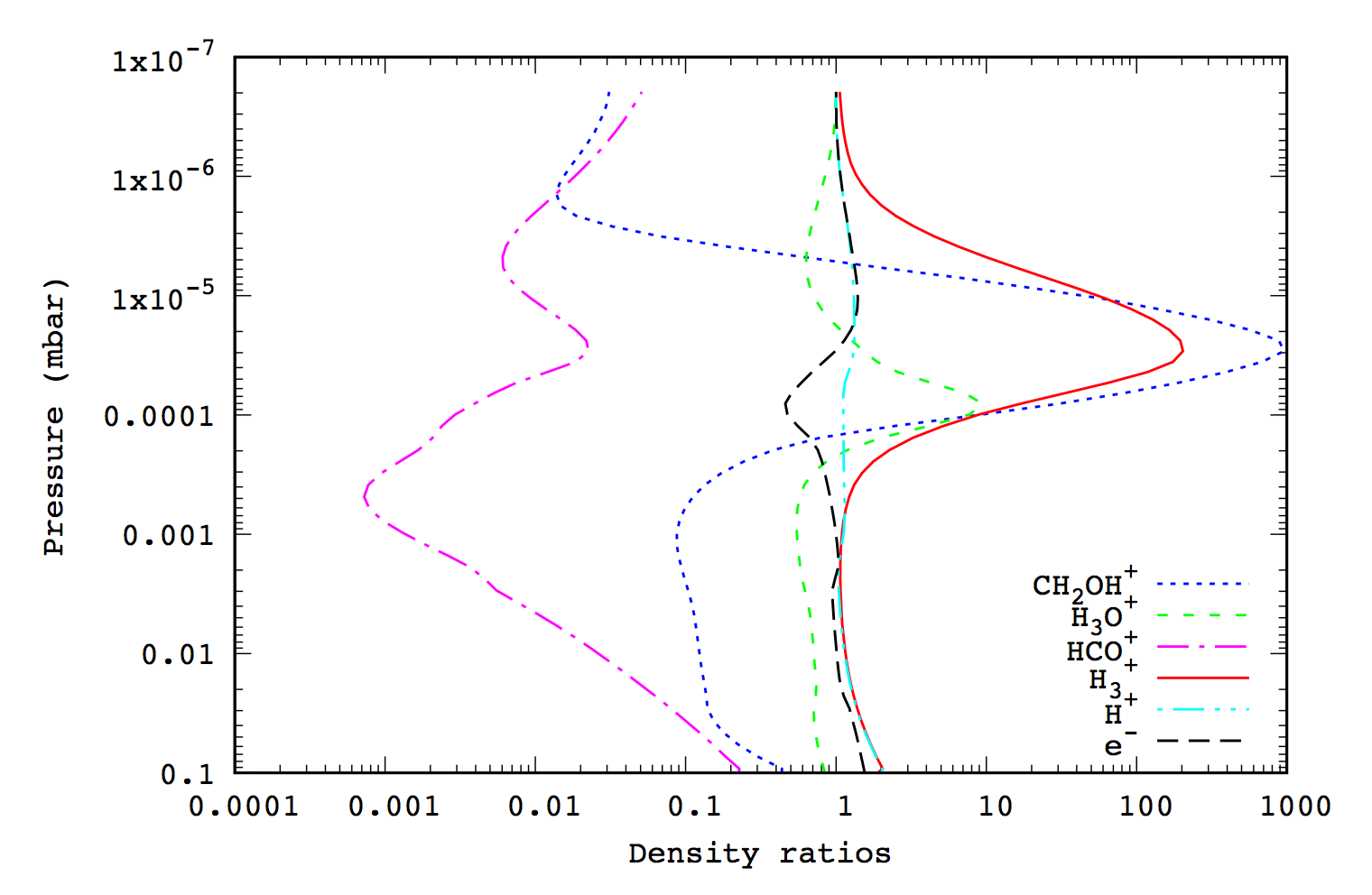} 
\includegraphics[width=1.0\columnwidth]{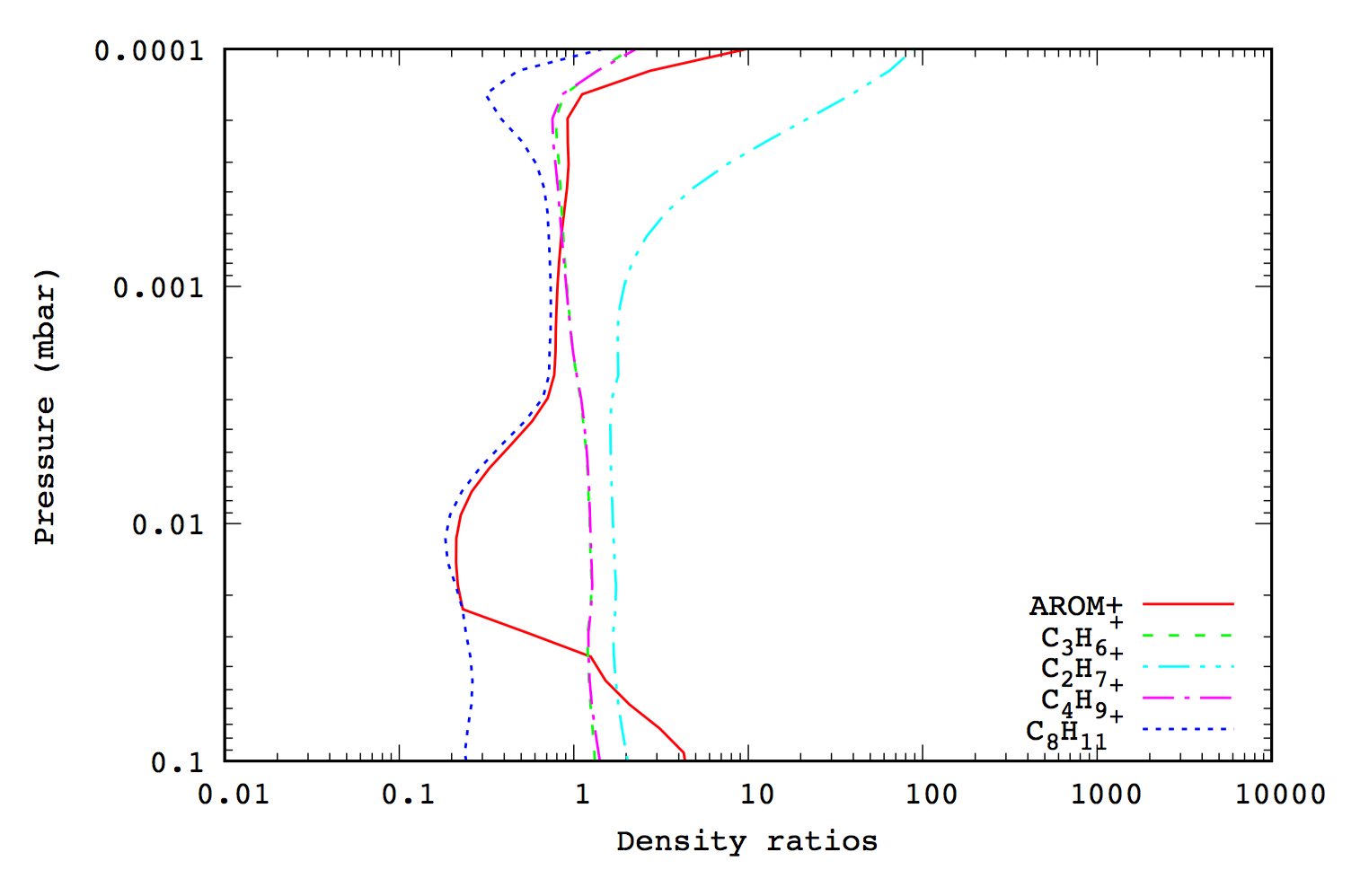} 
\caption{Density ratio profiles of the main ions for two different models. They correspond to the ratio of the density obtained with the model B1 (no external CO supply) over the density obtained with the model A2 (external CO supply).}
\label{fig:ratio}
\end{center}
\end{figure}

The influx rate of CO at the upper boundary of our model has an impact on the density profiles of some ions. Figure \ref{fig:ratio} shows the ratios of the densities of some ions obtained with a model with no influx of CO at the top of the atmosphere (model B1) and a model with a influx rate of CO equal to $2.0\times 10^8$ cm$^{-2}$s$^{-1}$ (model A2). H$^+$ is relatively unaffected by the influx rate of CO, whereas H$_3^+$, H$_3$O$^+$, CH$_2$OH$^+$, HCO$^+$ in the upper atmosphere and some heavy ions in the stratosphere are significantly affected with a density ratio of model B1 to A2 that can reach $10^3$ for CH$_2$OH$^+$ or $7\times 10^{-4}$ for HCO$^+$. When the CO abundance increases in the upper atmosphere, H$_3^+$ is more efficiently consumed (it reacts with CO) to produce HCO$^+$. Even if the proton affinities (PA) of H$_2$O and H$_2$CO favor the formation of H$_3$O$^+$ and CH$_2$OH$^+$ with respect to the one of HCO$^+$ (considering the PA of CO), the much larger CO abundance compared to H$_2$O and H$_2$CO clearly favors HCO$^+$ production. Consequently, future measurements of ion densities in the atmosphere (with a mass spectrometer for instance, see \citet{Mousis2018}) could give valuable constraints on the influx of CO and its origin.

CO absorbs UV light up to 165 nm (in our model) while its dissociation limit is around 108 nm. In the present model, we consider that the absorption between 108 nm and 165 nm leads to a complete attenuation of solar radiation in the atmosphere with a depth of penetration illustrated in Figure \ref{fig:penetration}. After this absorption, the excited CO should radiate a photon at similar wavelength more rapidly than being collisionally relaxed by surrounding molecules and atoms \citep{Field1983, Matsuda1991}. So, a part of these photons should be re-emitted to space, while a fraction of them should participate to the photolysis of other species. We tested a model considering that CO does not absorb between 108 nm and 165 nm and we found that the results (considering the effect on ions for instance) are quite similar to the model with no external supply of CO. Consequently, a detailed study on how the CO absorption contributes to the actinic flux in the range [108,165 nm] is required to better assess the importance of CO on the distributions of other species, and ions in particular, in the stratosphere of Neptune. 

\subsection{Uncertainty propagation study}
\label{section:UP}

Uncertainties on rate constants have noticeable effects on photochemical model results for many species (see for instance \cite{Dobrijevic2016} for the coupled ion-neutral model of Titan's atmosphere and \cite{Dobrijevic2010} for the neutral model of Neptune). In the present paper, we study the propagation of rate constant uncertainties in the case of the comet-impact hypothesis (model B2) and the ion-neutral model (model A2). 

\subsubsection{The cometary impact hypothesis}

\begin{figure}[htp]
\begin{center}
\includegraphics[width=0.8\columnwidth]{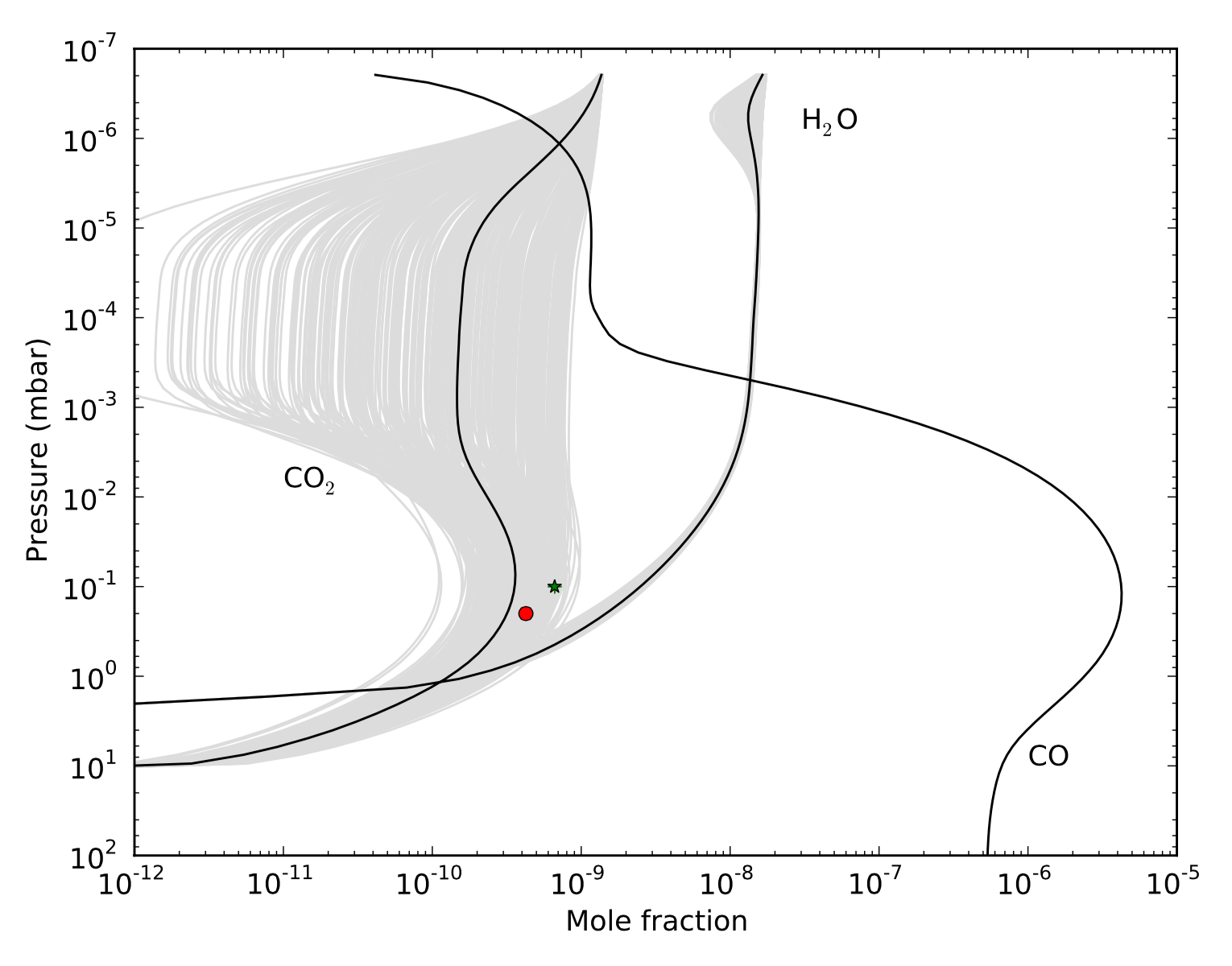} 
\includegraphics[width=0.8\columnwidth]{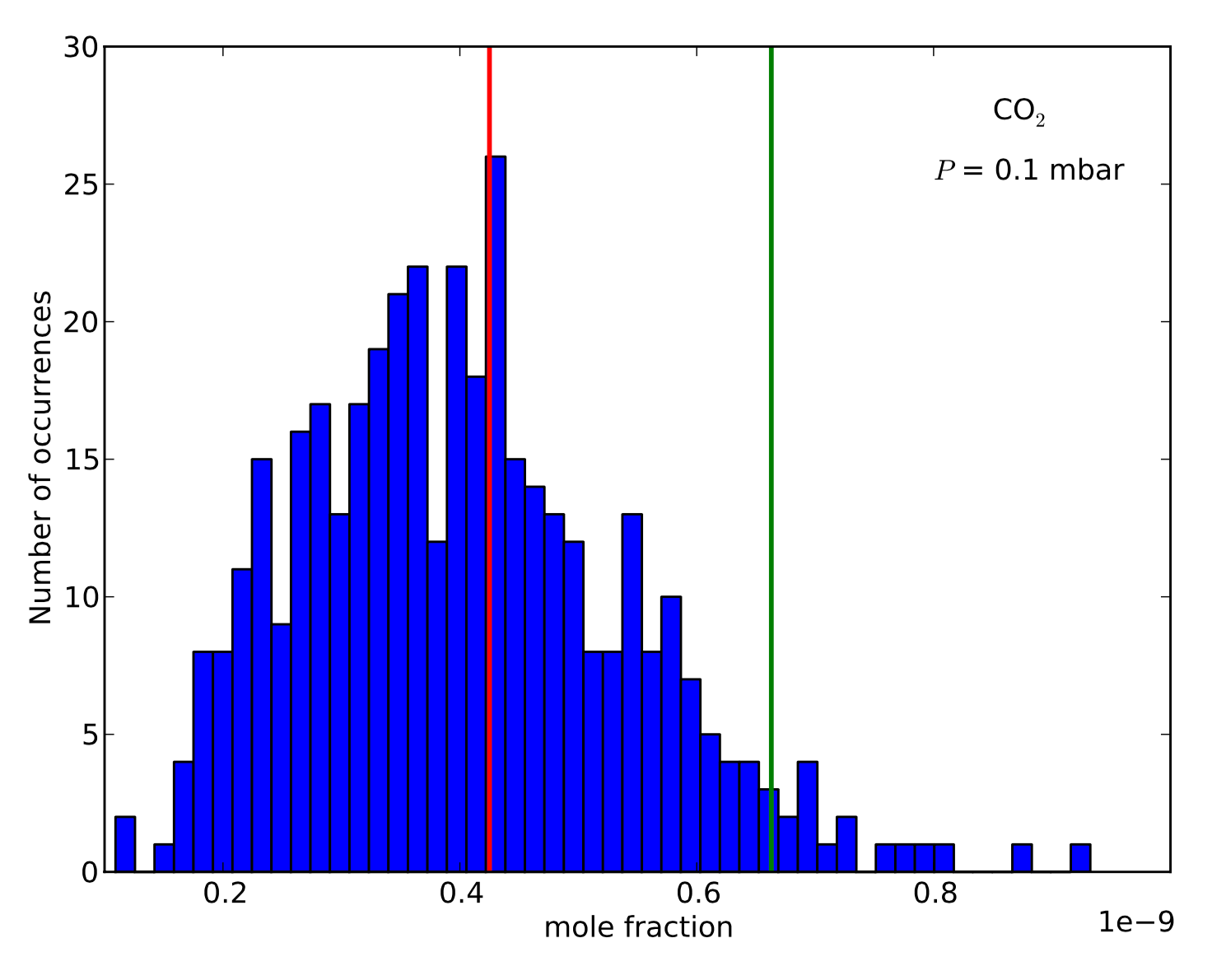} 
\caption{Top: Abundance profiles of CO, H$_2$O and CO$_2$ obtained after 400 runs of the Monte-Carlo procedure to account for the propagation of rate-constant uncertainties in our model. Thick black solid lines: initial profiles. Grey lines: Monte-Carlo profiles. Bottom: Distribution of CO$_2$ mole fractions at a pressure level of 0.1 mbar. Red and green values correspond to the abundances derived from observations by \citet{Feuchtgruber1997} and \citet{Meadows2008} respectively.}
\label{fig:MC_comet}
\end{center}
\end{figure}

In the case of the cometary impact hypothesis (model B2), we started the model with the initial profile of CO due to the impact (see Figure \ref{fig:comet}) and we stopped our model after 50 years for each Monte-Carlo run. For this integration time, we saw that the mole fraction profile of CO$_2$ is in quite good agreement with the observations in the nominal model. Results of our uncertainty propagation model are presented in Figure \ref{fig:MC_comet}. 

Uncertainties on the CO and H$_2$O vertical profiles are very small for such a short integration time. In contrast, uncertainties on the CO$_2$ vertical profile are relatively important, especially in the $10^{-3}$ - $10^{-5}$ mbar region for which, unfortunately, we do not have any observational constraints. Around the $10^{-1}$ mbar pressure level, the distribution of CO$_2$ mole fraction encompasses IR observations. The main consequences of these results are the following: (1) our CO$_2$ abundance profile obtained after 50 years of integration time is in agreement with observations, (2) taking into account uncertainties on model results, the date of the cometary impact that could have brought CO to the atmosphere of Neptune is quite imprecise. In particular, the results obtained by \citet{Moses2017} (impact 200 years ago) might be also considered to be in agreement with our results, given model uncertainties. 

An investigation of production and loss rates give us insight into the distribution of CO$_2$ profiles. The main reaction that contributes to the production of CO$_2$ throughout the atmosphere is  OH + CO $\rightarrow$ CO$_2$ + H (considering the integrated production rate) but around a pressure level of $10^{-4}$ mbar (600 km of altitude), an ionic chemical cycle involving HOCO$^+$ (through HOCO$^+$ + CH$_4$ $\rightarrow$ CH$_5^+$ + CO$_2$, H$_3^+$ + CO$_2$ $\rightarrow$ HOCO$^+$ + H$_2$, CH$_5^+$ + CO$_2$ $\rightarrow$ HOCO$^+$ + CH$_4$) and CO$_2^+$ (through CO$_2^+$ + H$_2$ $\rightarrow$ HOCO$^+$ + H, and the photoionisation of CO$_2$) controls the abundance of CO$_2$. The main loss process of CO$_2$ is the ionic reaction H$^+$ +  CO$_2$ $\rightarrow$ HCO$^+$ + O($^3$P) (with HCO$^+$ giving back CO). At the $10^{-4}$ mbar pressure level, both this reaction and photodissociation of CO$_2$ contribute to the loss of CO$_2$. Consequently, ionic chemistry has a strong influence on the abundance profile of CO$_2$. In particular, we found from a global sensitivity analysis (see the methodology of \citet{Dobrijevic2010}) that H$^+$ + H$_2$ $\rightarrow$ H$_3^+$ is a key reaction that has a strong influence on the uncertainties of CO$_2$ mole fractions at $10^{-4}$ mbar. 

\subsubsection{Major hydrocarbons}
\label{section:UP_hydroc}

\begin{figure}[htp]
\begin{center}
\includegraphics[width=0.8\columnwidth]{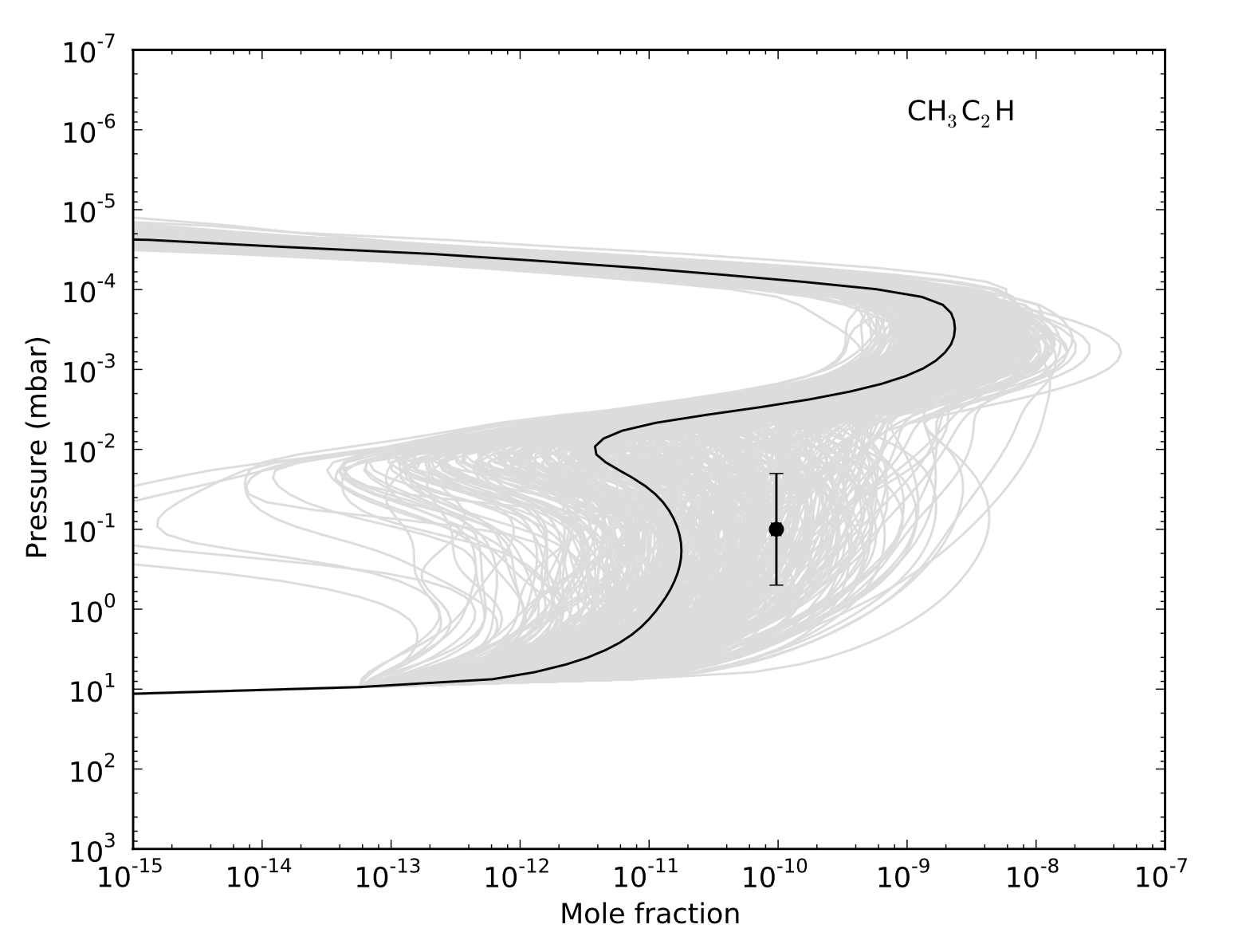} 
\includegraphics[width=0.8\columnwidth]{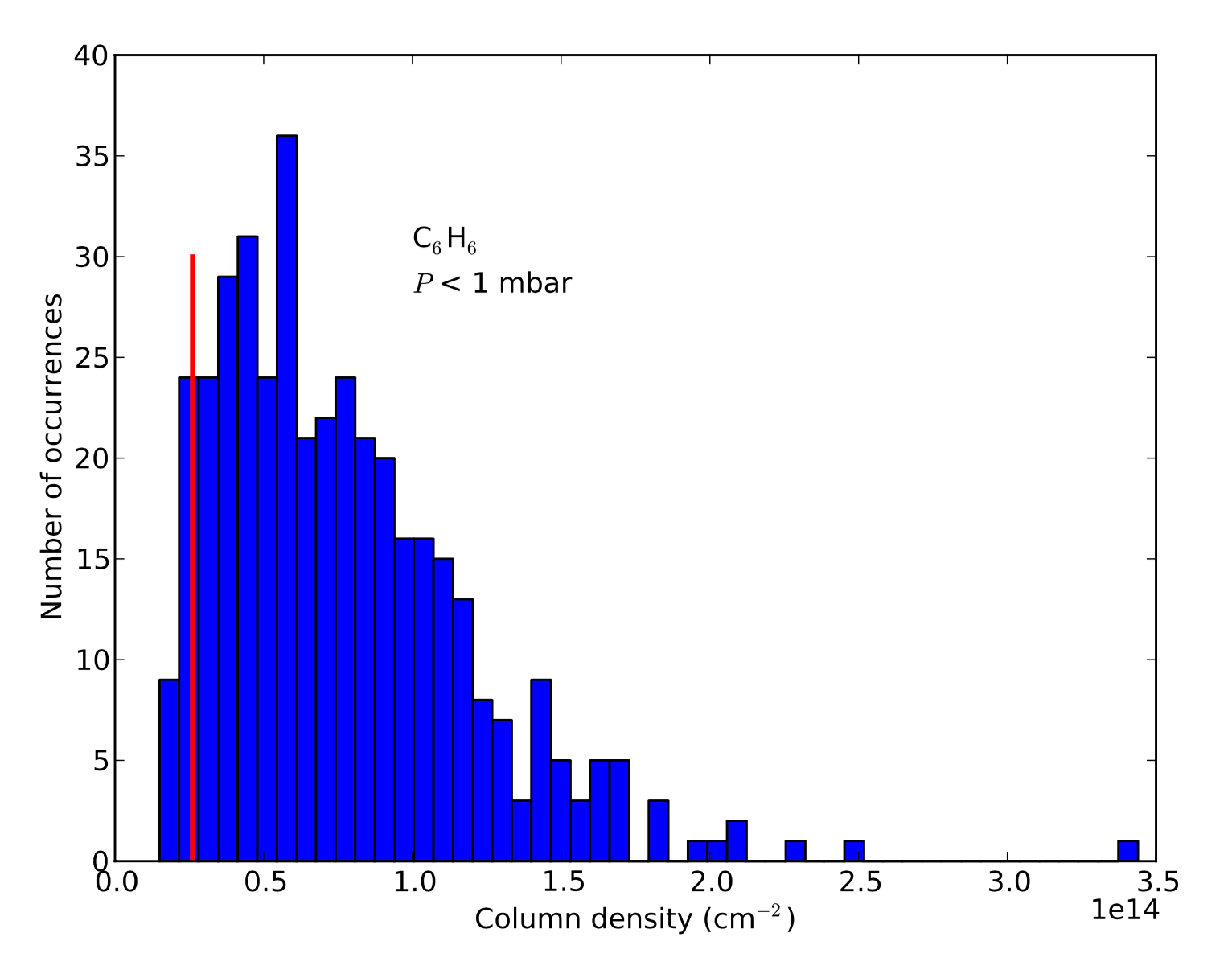} 
\caption{Top: Abundance profiles of CH$_3$C$_2$H obtained after 400 runs of the Monte-Carlo procedure to account for the propagation of rate constant uncertainties in our model. Thick black solid lines: initial profiles. Grey lines: Monte-Carlo profiles. Observation is given in black \citep{Meadows2008}. Bottom: Distribution of C$_6$H$_6$ column densities for pressure levels lower than 1 mbar. Red value corresponds to the upper limit of the column density derived by \citet{Bezard2001} from ISO observations.}
\label{fig:MC_hydroc}
\label{lastfig}
\end{center}
\end{figure}

We have also studied the propagation of rate constant uncertainties for hydrocarbons in the case of the ion-neutral model (model A2). We limit the presentation of our results to species with high uncertainties and observational constraints (CH$_3$C$_2$H and C$_6$H$_6$). 
We saw in Figure \ref{fig:C2} that our nominal model is not in agreement with the observation of \citet{Meadows2008} concerning CH$_3$C$_2$H. However, our uncertainty propagation results clearly shows in Figure \ref{fig:MC_hydroc} that the model abundance profile of this species is very uncertain, especially around 0.1 mbar. In fact, the abundance derived from the observation is in good agreement with our model considering the current knowledge of the rate constants of the reactions involved in the chemistry of CH$_3$C$_2$H.  The key reactions that are responsible for this dispersion are mainly:
\begin{center}
H + C$_2$H$_5$ $\rightarrow$ C$_2$H$_4$ + H$_2$ \\
$^1$CH$_2$ + H$_2$ $\rightarrow$ $^3$CH$_2$ + H$_2$ \\
$^1$CH$_2$ + H$_2$ $\rightarrow$ CH$_3$ + H \\
H + C$_2$H$_2$ $\rightarrow$ C$_2$H$_3$ \\
C$_6$H$_6^{**}$ $\rightarrow$ C$_6$H$_4$ + H$_2$ \\
\end{center}
These reactions have to be studied in priority to lower the uncertainty on model results concerning CH$_3$C$_2$H.

In Figure \ref{fig:MC_hydroc}, we have also depicted the Monte-Carlo distribution of C$_6$H$_6$ column densities to show the dispersion of results and to compare with the upper limit derived by \citet{Bezard2001} from ISO observations. It seems that our current model overpredicts the production of C$_6$H$_6$. It might be very useful to obtain new observational constraints on the abundance of this species.

Our results also reveal some difficulties interpreting photochemical results regarding possible seasonal variations \citep{Moses2018}. Depending on the rate constants used in the chemical scheme, it is not straightforward to be sure that models will give the same seasonal variations for all the chemical initial conditions. An uncertainty propagation study of a seasonal model seems to be mandatory to evaluate how confident we are on such theoretical variations.

\section{Conclusion}

The present study was motivated by several considerations: 

- The photochemistry of Titan's atmosphere reveals that benzene and other aromatics are efficiently produced by the coupling of ionic and neutral chemistries. So, we expect that these species should also be present in all giant planets. We tested the coupled chemical scheme used in \citet{Loison2018} (for Titan) in the case of Neptune, which might be considered as the least photochemically active planet of the Solar System due to its distance from the Sun. We have found that the coupling of ion and neutral chemistries produces aromatics in a relatively high abundance, several orders of magnitude greater than derived from neutral photochemical models. The photochemistry of hydrocarbons in Neptune's atmosphere is more complex than previously expected. We also predict that C$_3$H$_6$ might be relatively abundant and therefore detectable. If not, determination of an upper limit could be valuable for constraining the model.

- \citet{Guerlet2015} showed that the abundance of C$_6$H$_6$ in the auroral region of Saturn cannot be explained by neutral photochemical models and proposed that ion-neutral chemistry should be taken into account (see also \citet{Koskinen2016}). \citet{Hue2015} suggested the presence of large-scale stratospheric dynamics to explain the discrepancies between models and observations about mole fractions of C$_2$H$_2$ and C$_2$H$_6$ at high latitudes in the atmosphere of Saturn. Recently, \citet{Hue2018} tested this hypothesis for Jupiter and eventually proposed that the ion-neutral chemistry might be the best candidate to explain the distributions of C$_2$H$_2$ and C$_2$H$_6$ as a function of latitude in Jupiter's atmosphere. We have explored the effect of ionic chemistry on the abundances of C$_2$H$_6$ and C$_2$H$_2$ in the atmosphere of Neptune. We have found that it has no significant effect on C$_2$H$_6$ but decreases the abundance of C$_2$H$_2$ by a factor of about 1.5. We then expect that this effect on Jupiter's and Saturn's atmospheres might be more important.

- Several studies pointed out the necessity to include high-resolution cross sections of N$_2$ in photochemical models of Titan's atmosphere \citep{Lavvas2011}. \citet{Moses2017} discussed the putative effect of CO cross section resolution on the abundances of hydrocarbons and C$_2$H$_2$ in particular. Using high-resolution cross sections for several absorbing species in the ultraviolet (see \citet{Dobrijevic2016}), we do not see any significant difference in the abundance of C$_2$H$_2$ as a function of CO influx rate (in the neutral model). Cross sections with a even higher resolution than the ones we have used could be necessary to be more conclusive.

We also found some unexpected results:

- Two main ion layers are present in the atmosphere of Neptune with very different compositions. In the upper atmosphere ($10^{-5}$ mbar), the main ions are H$^+$, HCO$^+$, H$_2$O$^+$. In the stratosphere (around $10^{-3}$ mbar), there is a peak of hydrocarbon ions (like C$_2$H$_7^+$ and C$_5$H$_7^+$). However, according to \citet{Lyons1995}, the ablation of meteoroids could bring a substantial amount of metals in the stratosphere and Mg$^+$ might be the major contributor to the stratospheric peak. Updating the chemical scheme of metals could be valuable to confirm this point.

- We have shown that a large influx of oxygen species in the upper atmosphere of Neptune could have an effect on the concentration of several ions in the stratosphere. We have noticed however that a detailed study of the emission of CO in the stratosphere is required to better estimate its role in the actinic flux. According to previous models (see \citet{Moses2017} for instance), the origin of CO could be constrained by its abundance profile and by the abundance profile of CO$_2$. We have shown in the present study that an additional constraint can be brought by the determination of abundance profiles of several ions (like HCO$^+$). Consequently, a detection of these ions would give valuable information on the origin of CO.

- Based on the analysis of CO$_2$ observations, our nominal model favors a recent cometary impact (maybe less than 50 years) to explain the origin of CO, while \citet{Moses2017} proposed an impact with the same type of comet 200 years ago. Careful investigation of all the observations of CO and CO$_2$ using abundance profiles from photochemical models and radiative-transfer models are required to go further in this issue. We have also shown that ionic chemistry has a strong influence on the abundance profile of CO$_2$ in Neptune's atmosphere. Our uncertainty propagation study highlights strong uncertainties on the model results for CO$_2$ and a careful investigation of the main reactions involved in the ion-neutral chemistry of CO$_2$ is required to lower these uncertainties. Also, uncertainties in the eddy diffusion coefficient profile are currently important and have a strong effect on the model results regarding the evolution of CO, CO$_2$ and H$_2$O in the cometary impact hypothesis.

We conclude, from these results, that a better determination of the composition of Neptune's atmosphere (using adapted instruments on an orbiter and/or a mass spectrometer with an appropriate mass resolution) is essential to better understand the relative importance of ionic chemistry and the origin of CO and consequently the composition of Neptune's interior (in terms of oxygen content), which then can help us to decipher the various scenarii of Neptune's formation and evolution.

One major result of the present study is the demonstration that ionic chemistry substantially modifies the photochemistry of Neptune's atmosphere (and by analogy, the photochemistry of all giant planets). Since we included only photoionization in our model, the effect of ionization by magnetospheric electrons and galactic cosmic rays  should be investigated in future studies.

\newpage
\appendix

\textbf{Supplementary material}

\section{Integrated column rates}
\label{appendix:rates}

For each reaction included in the model, the integrated column rate (in cm$^{-2}$s$^{-1}$) scaled to the lower boundary ($z=-75$ km, $P= 9229.1$ mbar) and the mean altitude (in km) of the production are given.

\ack
Some parts of the computer time was provided by the "P\^ole Mod\'elisation HPC" facilities of the "Institut des Sciences Mol\'eculaires" (UMR 5255 CNRS, Universit\'e de Bordeaux) co-funded by the "r\'egion Nouvelle Aquitaine". 
We thank the "Programme National de Plan\'etologie" (PNP) of the "Institut National des Sciences de l'Univers" (INSU) for funding this work. This work was supported by the CNES.

\bibliographystyle{elsart-harv}


\label{lastpage}


\end{document}